\documentclass{aa}  

\usepackage{graphicx}
\usepackage{txfonts}
\usepackage{mathtools}
\usepackage{amsmath}
\usepackage{breqn}
\usepackage{float}
\usepackage{placeins}
\usepackage[normalem]{ulem}

\usepackage{natbib,twoopt}
\usepackage{xcolor}
\usepackage[breaklinks=true]{hyperref} 
\bibpunct{(}{)}{;}{a}{}{,}             
\makeatletter
  \newcommandtwoopt{\citeads}[3][][]{\href{http://adsabs.harvard.edu/abs/#3}%
    {\def\hyper@linkstart##1##2{}%
     \let\hyper@linkend\@empty\citealp[#1][#2]{#3}}}
  \newcommandtwoopt{\citepads}[3][][]{\href{http://adsabs.harvard.edu/abs/#3}%
    {\def\hyper@linkstart##1##2{}%
     \let\hyper@linkend\@empty\citep[#1][#2]{#3}}}
  \newcommandtwoopt{\citetads}[3][][]{\href{http://adsabs.harvard.edu/abs/#3}%
    {\def\hyper@linkstart##1##2{}%
     \let\hyper@linkend\@empty\citet[#1][#2]{#3}}}
  \newcommandtwoopt{\citeyearads}[3][][]%
    {\href{http://adsabs.harvard.edu/abs/#3}
    {\def\hyper@linkstart##1##2{}%
     \let\hyper@linkend\@empty\citeyear[#1][#2]{#3}}}
\makeatother

\newcommand{\spexxy}{\texttt{spexxy}}

\defcitealias{Vaz_2023}{MFIV}

\begin{document}

\title{The MUSE-Faint survey}

\subtitle{V. The binary fraction of Leo T}

\author{Daniel Vaz \inst{1}\fnmsep\inst{2}
  \and Jarle Brinchmann \inst{1}\fnmsep\inst{2}\fnmsep\inst{3}
  \and Sebastian Kamann \inst{4}
  \and Sara Saracino \inst{4}\fnmsep\inst{5}
  \and P. A. C. Cunha \inst{1}\fnmsep\inst{2}
  \and Mariana P. Júlio \inst{6}\fnmsep\inst{7}
}
\institute{Instituto de Astrofísica e Ciências do Espaço, Universidade do Porto, CAUP, Rua das Estrelas, 4150-762 Porto, Portugal \\ \email{Daniel.Vaz@astro.up.pt}
  \and 
  Departamento de Física e Astronomia, Faculdade de Ciências, Universidade do Porto, Rua do Campo Alegre 687, PT4169-007 Porto, Portugal
  \and
  Leiden Observatory, Leiden University, PO Box 9513, 2300 RA Leiden, The Netherlands
  \and 
  Astrophysics Research Institute, Liverpool John Moores University, 146 Brownlow Hill, Liverpool L3 5RF, UK
  \and
  INAF – Osservatorio Astrofisico di Arcetri, Largo E. Fermi 5, I-50125 Firenze, Italy
  \and
  Leibniz-Institute for Astrophysics Potsdam (AIP), An der Sternwarte 16, 14482 Potsdam, Germany
  \and
  Institut für Physik und Astronomie, Universität Potsdam, Karl-Liebknecht-Straße 24/25, D-14476 Potsdam, Germany
  \\
}

\date{Received ; accepted }
   
 
\abstract
{The Leo T dwarf galaxy, the faintest and least massive galaxy known to have recent star formation ($\leq 1~Gyr$), exhibits a high dynamical mass-to-light ratio ($100~\mathrm{\frac{M_\odot}{L_{V,\odot}}}$),  based on its stellar velocity dispersion ($7.07^{+1.29}_{-1.12}~\mathrm{km\ s^{-1}}$), indicating extreme dark matter dominance.}
{We present the first measurement of the binary fraction of Leo T using MUSE-Faint multi-epoch spectroscopy. We also determine the binary fraction for both young and old stellar populations separately and gain insights into binary properties in more metal-poor environments than the Milky Way or Magellanic Clouds. Finally, we investigate the potential impact of binaries on the inferred stellar velocity dispersion.}
{We employed a forward model methodology combining empirical scaling relations to predict stellar velocity variations and a constrained binary distribution from the literature. To estimate the close binary fraction, we limited the maximum semi-major axis ($a < 10~\mathrm{au}$) and repeated the analysis with a semi-amplitude threshold ($\geq~10~ \mathrm{km\ s^{-1}}$) to check the impact on the inferred stellar velocity dispersion.}
{The overall binary fraction of Leo T is estimated to be $55^{+40}_{-9} \%$, consistent with similar systems. The close binary fraction ($a < 10~\mathrm{au}$) is $30^{+34}_{-9} \%$, which is aligned with low-metallicity environments. 
We found a lower binary fraction for the older stellar population ($15^{+43}_{-15} \%$) when compared to the younger population ($35^{+40}_{-6} \%$). 
Finally, we found no significant inflation of the velocity dispersion estimate due to binary motions when compared to the dispersion inferred from the co-added spectra. This suggests that the co-added spectra effectively provide period-averaged velocities of the stars, thus mitigating the impact of binaries on the overall velocity dispersion measurement.
}
{}


\maketitle
%
\section{Introduction}
A multitude of exceedingly faint dwarf galaxies have been detected near the Milky Way \citep{Willman_2010, Simon_2019}. Termed ultra-faint dwarfs (UFDs), they demarcate the low-mass limit for galaxy formation \citep{Belokurov_2013}, and they are the oldest, most metal-poor, and least chemically evolved stellar systems known \citep{Simon_2019}. These small-scale systems are dispersion supported \citep{Wheeler_2017}, and therefore a precise measurement of the velocity dispersion of their stellar components is crucial to accurately determining their dynamical masses \citep{Mateo_1993, Wolf_2010}. 
From recent measurements of the velocity dispersion of these systems, it has been shown that they boast the highest mass-to-light ratio among all known galaxies, granting them the status of the most dark matter dominated systems known in the cosmos \citep{Simon_2007, Amorisco_2011, Bullock_2017}.
Naturally, these findings have drawn substantial attention to UFDs, prompting various efforts to obtain the most precise measurements of their velocity dispersion \citep[see][ for a review]{Battaglia_2022}. Since typical velocity dispersions for UFDs are $\sigma < 10~\mathrm{km}\,\mathrm{s}^{-1}$, one of the main challenges in achieving this precision lies in determining the contamination and impact of binary stars on these measurements, given their potential to inflate the observed velocity dispersions of these galaxies \citep[e.g.][]{Martinez_2011, Minor_2013}.

For classical dwarf spheroidal galaxies, in the cases where the measured velocity dispersion falls between $4~\mathrm{and}~10\,\mathrm{km\ s^{-1}}$, it has been suggested that the likelihood of the measurement being inflated by more than 30\% due to the orbital motion of binary stars is very low \citep{Minor_2010}. Extending this premise to UFDs, it is assumed that binary stars do not inflate the observed velocity dispersion, although this assumption is not inherently evident. In theory, one might anticipate a greater influence of binaries in UFDs due to their limited star count and smaller intrinsic velocity dispersions. Additionally, concern for binary stars in UFDs is amplified by recent findings suggesting substantially higher binary star fractions in systems with lower metallicities \citep{Raghavan_2010, Badenes_2018, Moe_2019}. 

Investigations of UFDs with multi-epoch spectroscopic measurements have indicated that neglecting stars exhibiting noticeable velocity variations between epochs reduces the measured value of the velocity dispersion \citep[e.g.][]{Ji_2016J, Kirby_2017}.
Evidence suggests that if the intrinsic velocity dispersion of the system is below $4~\mathrm{km/s}$, the measurement could be inflated by more than 100\% due to binary star motion \citep{Spencer_2017, Spencer_2018, Koposov_2011}. Moreover, \cite{McConnachie_2010} demonstrated a probability exceeding 20\% that the dispersions of several UFDs hover around $\sim 0.2~\mathrm{km\ s^{-1}}$, similar to globular clusters, with the presence of binaries elevating the measured velocity dispersions to the values measured today.

Recent studies have sought to quantify the influence of binary stars on the estimation of dynamical masses in dwarf galaxies \citep[e.g.][]{Pianta_2022}. However, the small sample of stars in these systems and the lack of multi-epoch observations that would reveal velocity changes and expose hidden binaries make it hard to measure their binary fraction and period distribution with confidence. Hence, in only a few cases do the available spectroscopic data provide sufficient information to constrain the binary fraction \citep[e.g.][]{Minor_2019, Arroyo_Polonio_2023}. With regard to period distributions, the task becomes even more challenging, as it is exceedingly rare to find examples within the existing literature \citep[e.g.][where orbital parameters for a Triangulum II binary star member are found]{Buttry_2022}.
Alternative methodologies have been attempted, such as in \cite{Safarzadeh_2022}, where the authors explore wide binaries using nearest-neighbour statistics. 

Much is still unknown in this field, and we need to study more dwarf galaxies to gain a full understanding of the binary stars in these systems, estimate their binary fractions and distributions of orbital parameters, and determine how binaries affect the velocity dispersion that we are measuring.
Our objective is to contribute towards this effort with an additional assessment of the binary fraction within a faint dwarf and to develop a methodology that we will apply to other dwarfs in the future.
We present the first measurement of the binary fraction of the Leo T dwarf galaxy, utilising multi-epoch spectroscopic observations via the Multi-Unit Spectroscopic Explorer \citep[MUSE;][]{baconMUSESecondgenerationVLT2010} integral field spectrograph. This is part of the MUSE-Faint survey of UFDs \citep{Zoutendijk_2020}. Leo T has previously undergone analysis through the MUSE-Faint survey, with the results documented in \cite{Vaz_2023}, referred to as \citetalias{Vaz_2023} from this point on. A pertinent result from this previous study is the identification of two dynamically distinct populations within Leo T: a young population younger than 1 Gyr and an old population older than 5 Gyr.

This paper is structured as follows: Section~\ref{sec:observations} outlines the data we used, providing a summary of our observations and an overview of the data reduction procedures. In Section~\ref{sec:data_analysis}, we elaborate on the methods employed, detailing the procedural steps applied for data analysis. In Section~\ref{sec:FM} we present the forward modelling methodology used to estimate the binary fraction of Leo T.
The results of our analysis and a discussion are presented in Section~\ref{sec:results}.
We draw our conclusions and summarise the key takeaways from this paper in Section~\ref{sec:conclusions}.

\section{Observations and data reduction}
\label{sec:observations}

In this section, we provide an overview of the data and the process undertaken for analysing the multi-epoch data of Leo T. We summarise the observations,  the data reduction process, and detail the method employed to obtain the multi-epoch stellar spectra. Table~\ref{tab:LeoTP} presents relevant Leo T properties from \citetalias{Vaz_2023}.

\begin{table}[]
\centering
\caption{Relevant Leo T properties.}
\begin{tabular}{lccrr}
\hline\hline
\multicolumn{1}{c}{Property}                                     & \multicolumn{1}{c}{Value}                    \\ \hline
$\alpha_{J2000}$                                     & $09\ 34\ 53.4$             \\
$\delta_{J2000}$                                 & $+17\ 03\ 05$             \\
Distance $\mathrm{[kpc]^{(1)}}$                                 & $409^{+29}_{-27}$             \\
Luminosity $\mathrm{[L_{\odot,V}]^{(2)}}$                       & $1.4 \times 10^5$                  \\
3D $r_{1/2}~\mathrm{[pc]^{(2)}}$                                         & $152 \pm 21$                       \\
2D $r_{1/2}~\mathrm{[pc]^{(2)}}$                                   & $115 \pm 17$                       \\
$v_\star$ $\mathrm{[km\ s^{-1}]^{(3)}}$                       & $39.39^{+1.32}_{-1.29}$                       \\
$\sigma_\star$ $\mathrm{[km\ s^{-1}]^{(3)}}$                  & $7.07^{+1.29}_{-1.12}$                      \\
$\sigma_{\star young}$ $\mathrm{[km\ s^{-1}]^{(3)}}$                  & $2.31^{+2.68}_{-1.65}$                      \\
$\sigma_{\star old}$ $\mathrm{[km\ s^{-1}]^{(3)}}$                  & $8.14^{+1.66}_{-1.38}$                      \\
$\mathrm{[Fe/H]}^{(3,4)}$                                             & $\sim -1.6$                        \\
\hline
\end{tabular}
\tablebib{$^1$\citet{Clementini_2012}; $^2$\citet{de_Jong_2008};
  $^3$\citet{Vaz_2023};
  $^4$\citet{Weisz_2012}.}
	\label{tab:LeoTP}
\end{table}

\subsection{Observations}

The central region of Leo T was observed as part of the MUSE-Faint project \citep{Zoutendijk_2020}, a survey dedicated to UFD galaxies (PI Brinchmann). MUSE, a large-field medium-spectral-resolution integrated field spectrograph, is housed at Unit Telescope 4 of the Very Large Telescope (VLT).
For Leo T observations, we used the Wide Field Mode with ground-layer adaptive optics (WFM-AO), presenting a $1 \times 1 \; \mathrm{arcmin^2}$ field of view segregated into 24 slices, each directed to an individual integral field unit (IFU). This configuration ensures a spatial sampling of $ 0.2\;\mathrm{arcsec\,pixel^{-1}}$ and a wavelength sampling of $1.25$ \AA\, $\mathrm{pixel^{-1}}$. The nominal wavelength coverage spans $4700-9350$ \AA, with a notch filter specifically excluding the range 5807--5963 \AA\ to prevent contamination from sodium lasers used in AO mode.

We treated each observation block (OB) as an independent epoch to preserve sensitivity to velocity variations; exposures within an OB are co-added, but exposures from different OBs are never combined. The final dataset comprises five OBs that satisfied the observatory quality control grade “A” (optimal conditions). Four OBs contain three 900-s exposures each, and one additional OB obtained on 22–24 December 2022 consists of six 829-s exposures. An additional OB graded “B” due to poor seeing (FWHM $\approx 0.96$ arcsec at $7000$ \AA) was processed but yielded no spectra; consequently, it provides no usable velocity measurements and is excluded from the analysis.
In Table~\ref{tab:obs} we include details of the OBs used, including the exposure time, and full
width at half maximum (FWHM) at 7000 \AA.

\begin{table}[ht]
\footnotesize
\centering
\caption{List of the MUSE observations of Leo T, with information of the exposure time, and full width at half maximum (FWHM) at 7000 \AA. }
\begin{tabular}{lcccrr}
    \hline\hline
    \multicolumn{1}{c}{Date} & \multicolumn{1}{c}{Exp. Time}  & \multicolumn{1}{c}{FWHM @ 7000\AA} \\ 
                             & \multicolumn{1}{c}{[min.]}    & \multicolumn{1}{c}{[arcsec]} \\
    \hline
    2018-02-14    & 45                    & 0.52 \\
    2018-03-17    & 45                    & 0.53 \\
    2018-04-16/17 & 45                    & 0.69 \\
    2019-04-09    & 45                    & 0.60  \\
    2022-12-22/23/24 & 82.9               & 0.55  \\
    \hline
    \end{tabular}
\label{tab:obs}
\tablefoot{The first four observation blocks correspond to three exposures of 15 minutes, while the last observation block are six exposures of 829 seconds each.}
\end{table}

\subsection{Data reduction}

Data reduction followed the process described in \citetalias{Vaz_2023}.  We processed the data using the ESO Recipe Execution Tool (EsoRex; version 3.13), using the MUSE Data Reduction Software (DRS; version 2.8.1; \citealt{weilbacher2020data}). 
The reduction process adhered to standard procedures. 
This data processing methodology resulted in the creation of a data cube per OB and hence a data cube per epoch of observation. 

To extract stellar spectra from the five data cubes, our methodology again followed that of \citetalias{Vaz_2023}. In summary, we used PampelMuse \citep{Kamann_2012}, specifically optimised for extracting stars from integral-field spectroscopic observations, especially in densely populated stellar fields. This software needs a high-fidelity source catalogue, typically sourced from data such as HST, to identify and pinpoint sources within the data cube. We used the identical master list of sources of \citetalias{Vaz_2023}, derived from public HST ACS data of Leo T.\footnote{HST Proposals 12914, Principal Investigator Tuan Do
and 14224, Principal Investigator Carme Gallart}

Starting with our star selection from \citetalias{Vaz_2023}, we tried to locate the same stars within the updated data cubes. Successfully, we managed to identify around 252 stars per data cube, remaining consistent with our previous count. Although this count might fluctuate slightly per data cube, the count remains consistent within $\pm 3$ stars. This does not affect the expected final sample, as we also make some quality cuts, as we discuss next. 

\section{Data analyses}
\label{sec:data_analysis}

Despite our existing knowledge regarding the likely membership status of specific stars within Leo T, subtle differences in the new dataset, combined with our refined analysis, may now lead to the inclusion of some stars previously rejected by our quality criteria and the exclusion of others that had formerly satisfied them.
In this new analysis, we expect a lower signal-to-noise ratio (S/N) because the data cubes are not combined, so each spectrum effectively has a shorter exposure time. This might lead to some stars previously identified as members of Leo T now not being suitable for analysis. 
The previous combination of exposures taken at different times might impact certain stellar spectra, potentially causing perturbations that could affect processes such as velocity fitting \citep[e.g.][]{Badry_2017}. This would not happen this time, allowing some stars that were not possible to analyse before to be present in the new sample.

To ensure a robust selection of the stellar population of Leo T from this new dataset, we replicate the same spectrum analysis conducted in \citetalias{Vaz_2023}, maintaining the criteria. 
Moreover, we introduce a new quality criterion that only considers stars with more than one velocity measurement, as it is essential for our study to have multiple velocity measurements per star.  

\subsection{Measuring stellar properties}

To be consistent with our previous analysis, we adopt the parameters for Leo T from \citetalias{Vaz_2023}. In the following, we describe our methodology and the steps that we took in order to measure the stellar properties needed for our subsequent analysis. 

\subsubsection{Measuring velocities}
\label{sec:spexxy_vel}

We adopted the same methodology as in \citetalias{Vaz_2023} and use the \spexxy\ full-spectrum fitting code \citep{gup-87} to estimate the physical parameters from which we extracted the line-of-sight velocity. This code operates by conducting an interpolation across a grid of PHOENIX model spectra \citep{Husser_2013}. For more details, see \citetalias{Vaz_2023}.

\subsubsection{Measuring stellar masses}
\label{sec:spexxy_masses}

We estimated the effective temperatures, log g, and stellar masses through fits of isochrones to the observed photometry for the stars.
We used a \spexxy\ tool called \texttt{isochrone}, with the command \texttt{apply}, which applies a designated isochrone to all stars within a specified photometry file. This process derives the effective temperatures, surface gravities, and current masses for each star.

\begin{center}
\begin{figure}[]
\resizebox{\hsize}{!}{\includegraphics[scale=1]{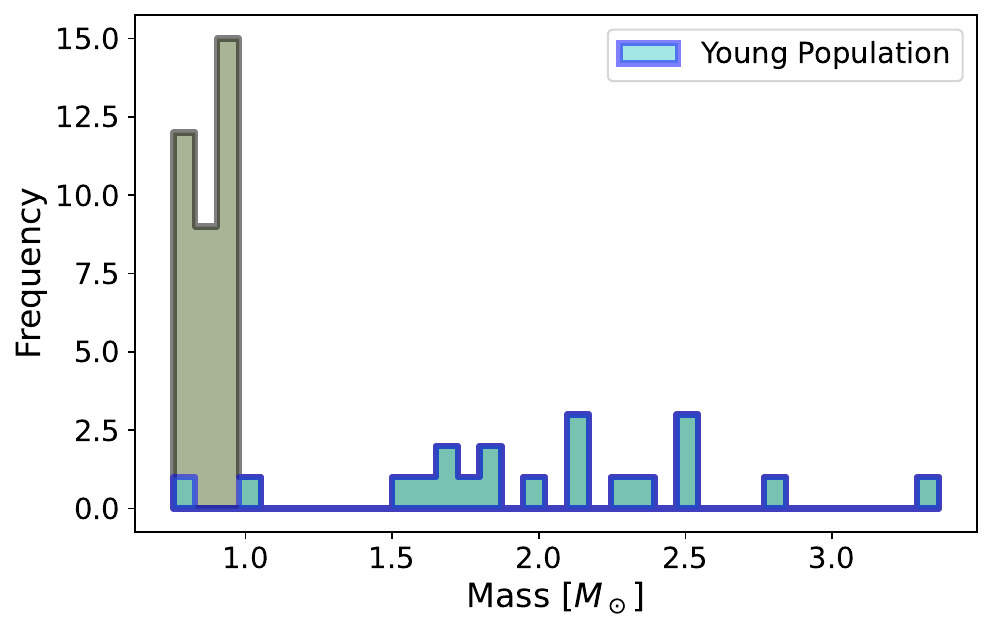}}
        \caption{Distributions of the estimated masses using \spexxy\ tool. The fit was made using the nearest-neighbour method. The full sample is shown, with identification of the young population in blue.} 
        \label{fig:mass_hist}
\end{figure}
\end{center}

We used the same isochrones as in \citetalias{Vaz_2023}, derived from PARSEC stellar tracks and isochrones \citep{Bressan_2012}. Specifically, we assume a fixed metallicity of $\mathrm{[M/H}]=-1.6$ and adopt an interstellar extinction of $A_V = 0.1$ magnitudes, which aligns closely with the value of $A_{V_{SFD}} = 0.0959$ that we get from \cite{SFD1998}. We generate 10 isochrones equally spaced between 0.1 and 1 Gyr, and 7 isochrones equally spaced between 5 and 11 Gyr, which serve as input to the \spexxy. We estimate the stellar masses using the nearest-neighbour method.
Figure~\ref{fig:mass_hist} displays a histogram of the estimated masses.

\subsection{Selecting members}\label{sec:mem}

We initially selected spectra with a median S/N$>3$, per $1.5~\AA$. Given our five spectra per star (corresponding to five epochs) with varying median S/N, this cut was applied per epoch, resulting in a non-uniform selection where individual epochs for a star might or might not meet this S/N threshold.
We evaluated the success of our fitting process for each spectrum and restricted our analysis to stars with more than one successful velocity measurement, resulting in a final sample of 55 stars. We note that, as a result of these selection criteria, the minimum S/N of the spectra in our final sample is always $>4$, ensuring the reliability of the velocity measurements.

Of our 55 stars, 32 have velocity measurements from all five epochs, 9 have measurements from three epochs, and the remaining 14 have measurements from only two epochs. The data quality varied across epochs, with the last epoch having the longest exposure time and thus the highest S/N. The second epoch also provided notably high-quality spectra. Consequently, all 55 stars in our final sample have measurements from these two highest-quality epochs. The third epoch offered the next best data quality, meaning that the 9 stars with at least three velocity measurements also include data from this epoch.
This pattern is reflected in the typical velocity uncertainties of each epoch: for stars with five epochs, the best epochs yield uncertainties of $\lesssim 10$ km s$^{-1}$, while the worst epochs are closer to $\sim$20 km s$^{-1}$. As a result, most stars have two epochs with uncertainties of about 15 km s$^{-1}$ and another three epochs with slightly higher values. For stars covered in fewer epochs, the typical uncertainties remain below $\sim$20 km s$^{-1}$. A histogram illustrating these distributions for each epoch, as well as the combined sample, is presented in Appendix~\ref{app:violin} (Figure~\ref{fig:uncertainty_hists}).

\begin{figure*}[t]
\centering
\includegraphics[scale=1]{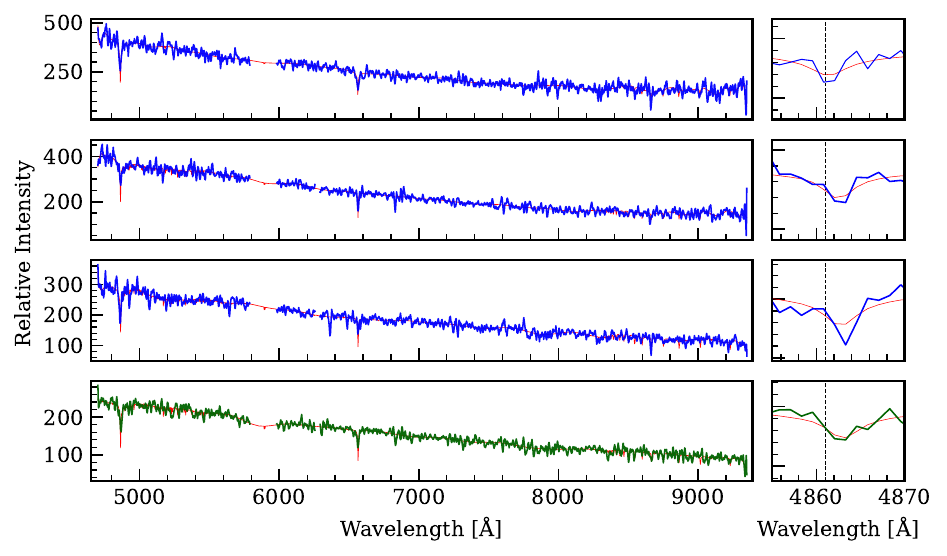}
\caption{Examples of spectra along with their corresponding \spexxy\ best fit utilising three different epochs for a star likely to be in a binary system. The lower panel displays the combined spectrum from all five epochs, plotted in green. Given that this star is one of the most probable binaries in our sample, we provide a zoomed-in view of the $H\beta$ line, with a dotted black line indicating the expected line centre in the rest frame to illustrate the shift in velocity from epoch to epoch. This shift between epochs affects the co-added spectrum, where the line is strongly diminished as a result of multiple velocity components blending together.}
        \label{fig:spec_examples}
\end{figure*}

Of the 55 stars in this sample, 44 overlap with the sample described in \citetalias{Vaz_2023}. The remaining 11 stars were not part of the previous sample due to issues encountered during spectral fitting. The reasons for these fitting failures vary, but in most cases they are linked to offsets between epochs. For example, when individual spectra with different line-of-sight velocities are co-added without correcting for these shifts, the absorption lines are broadened or diminished, which can prevent a reliable fit. In one clear example, the combined spectrum exhibits double-peaked lines, strongly suggesting a binary system. Figure~\ref{fig:spec_examples} illustrates this case with spectra from three epochs, and a zoom-in on the $H\beta$ line reveals a clear shift from epoch to epoch. Of the 11 stars that failed in MFIV and are included here, 4 are now identified as likely binary candidates, as shown below. Another issue arises in the case of bright stars where the velocity measurements across epochs are mutually consistent, but the metallicity determinations vary substantially; here, the fitting failure seems to stem from inconsistent metallicity estimates rather than binary motion. Finally, some of the failed cases are affected by two or more low-S/N epochs, making it impossible to establish whether the poor fits are caused by true velocity variations or simply by limited data quality.

\subsection{Identifying populations}\label{sec:pop}

Due to the new possible members of Leo T within this new sample, and knowing that Leo T possesses multiple stellar populations, we repeat here a photometric analysis, using the same methods as in \citetalias{Vaz_2023}. We used F606W and F814W photometry of the public HST/ACS data, the same that was used to create a master list of sources, fit for the stellar masses, and also used in \citetalias{Vaz_2023}. We constructed a colour-magnitude diagram and juxtaposed it with isochrones derived from PARSEC stellar tracks and isochrones, similar to the one employed in the stellar mass fitting. The nearest-neighbour method was applied to allocate an age to each star. Instead of utilising individual ages, we categorised the stars into two groups: one that encompasses stars with ages of $\leq 1$ Gyr and another group comprising stars aged $\geq 5$ Gyr.

\begin{figure}[]
\resizebox{\hsize}{!} {\includegraphics[scale=1]{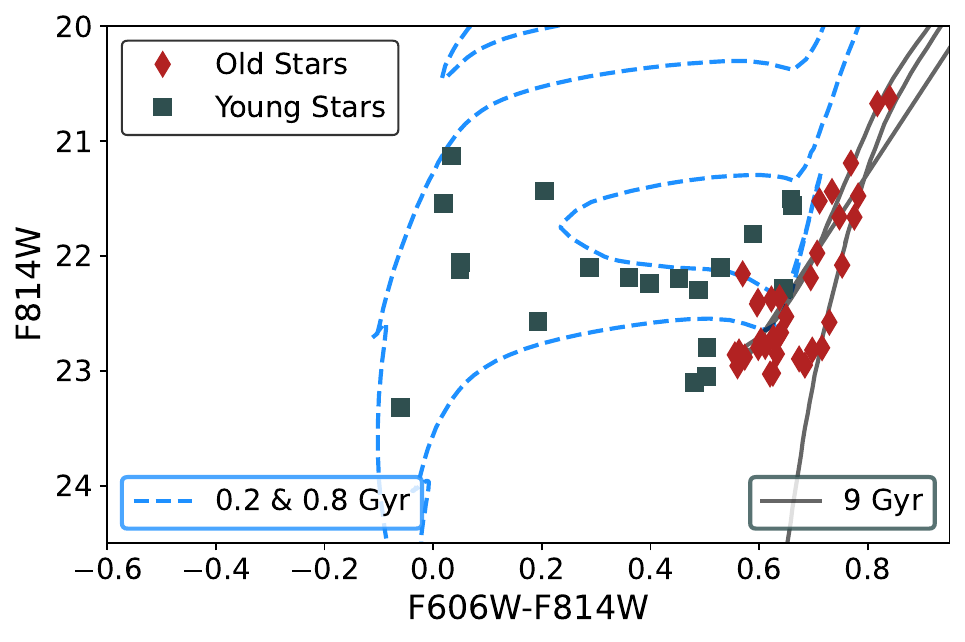}}
\caption{Colour-magnitude diagram of the 55 stars plotted against PARSEC isochrones drawn for constant $[\mathrm{Fe/H}] = -1.6$ and variable age. The magnitudes given are on the Vega Magnitude System.
  The region spanned by $0.1-1$ Gyr isochrones is represented by two isochrones shown for 0.2 and 0.8 Gyr. We also show a 9 Gyr isochrone for illustration of the old region spanned by $>5\,\mathrm{Gyr}$. 
  The stars that were found to be consistent with the younger isochrones are shown as dark blue squares while the stars consistent with the older isochrones are shown as red diamonds.}
        \label{fig:color_plot}
\end{figure}

Figure~\ref{fig:color_plot} displays the colour-magnitude diagram presenting the 55 stars plotted alongside the PARSEC isochrones. The stars are differentiated by different symbols and colours, each denoting the assigned age group: red diamonds represent older stars aged $\geq 5$ Gyr, while blue squares indicate younger stars aged $\leq 1$ Gyr. Three representative isochrones are depicted for 0.2, 1.0, and 9 Gyr.

All 55 stars are in line with the isochrones, affirming their suitability for further analysis in this study. Among the stars, 35 are identified as members of the older sample, while the remaining 20 belong to the younger sample.

\section{Forward modelling}
\label{sec:FM}

In order to determine the binary fraction of the observed stellar population, we relied on a forward model methodology. The strategy we used is based on the methodology presented in \cite{Giesers_2019} and also on the methodology presented in \cite{Arroyo_Polonio_2023}, which in turn is based on the strategy of \cite{Spencer_2018}.

Our method relies on multi-epoch observations of stars and detects radial velocity variations through the use of the reduced-$\chi^2$ statistics (we denote this simply as $\chi^2$). We computed $\chi^2$ as 
\begin{equation}\label{eq:chi2}
\chi_i ^2=\frac{1}{\kappa} \sum_{i}^m \left( \frac{v_i-\bar{v}}{\sigma_i} \right)^2,
\end{equation}
where  $v_i$ is a single velocity measurement, $\sigma_i$ is the corresponding velocity measurement error,
and $\kappa = m-1$ is the number of degrees of freedom, where $m$ is the number of observations. We fit $\bar{v}$ for each star, choosing the value that minimises the $\chi^2$ value, thus the mean velocity weighted by $1/\sigma^2$. By doing so, we are removing the dependence on the velocity component related to the systemic motion of the star, which we call $v_0$ in the remainder of this text.

In order to quantify the binary fraction of the observed stars, we simulate mock datasets with the same characteristics as the observed dataset. We emulate the number of stars, the number of velocity measurements per star, the uncertainty of each measurement, the time of the measurements, and the mass of each star. 
Each dataset is created 101 times, all with different binary fractions ($f$), ranging from 0 to 100\%, in steps of 1\%. In order to cover all the parameter space for all the variables that we consider, we generate 10,000 mock samples per binary fraction. 

\subsection{The model}

In order to create these datasets, we proceed as follows for each fraction $f$:
We randomly classified each observed star as either a binary or a non-binary, based on $f$. We  then assigned a mock $v_{los}$ to each observed star for every epoch, consistent with its classification as binary or non-binary.

For non-binaries, we simplify $v_{los}$ to two components, the systematic velocity ($v_0$), which includes the peculiar velocity of Leo T and the individual star's velocity within the galaxy, and the measurement uncertainty $\delta$. Consequently, we define the mock $v_{los}$ for a non-binary as
\begin{equation}
    v_{nb} = v_0 +\delta.
\end{equation}

For binaries, $v_{los}$ comprises three components: the aforementioned $v_0$ and $\delta$, and the radial velocity arising from the binary orbit, $v_b$. However, the spectral resolution of MUSE limits the measured radial velocity amplitude, an effect influenced by the flux ratio ($r =\frac{f_2}{f_1}$, where $f_1$ is the brighter component) of the two stars in the binary system \citep{Giesers_2019}. Specifically, this measured radial velocity is linearly damped by the flux ratio. Therefore, we define the mock $v_{los}$ for a binary as
\begin{equation}
    v_{b} = v_0 + (1-r)~v_b + \delta.
\end{equation}
We note that in our mock data the flux ratio is typically small, with the large majority of systems having $r < 0.1$, making the damping effect minimal.

The simulation of the binary radial velocity $v_b$ is detailed in Section~\ref{sec:BinDists}, and depends on the adopted binary population distributions, the mass of the observed star, and the temporal variations across epochs. When flux values for the simulated stars are required, we perform the inverse of the mass estimation process described in Section~\ref{sec:spexxy_masses}. 

The systemic velocity $v_0$ is randomly drawn for each star and is held constant across all epochs.  We assume that $v_0$ follows a Gaussian distribution with the parameters derived for the target galaxy. Specifically, for Leo T, we adopt the values from \citetalias{Vaz_2023}, $v_{mean} = 39.39~\mathrm{km\,s^{-1}}$ and $\sigma_v = 7.07~\mathrm{km\,s^{-1}}$. Although the calculation of our $\chi^2$ values is independent of these systemic velocity assignments, we perform this step to generate a more realistic representation of the observed data, facilitating potential future analyses and tests using the same mock dataset.

The measurement uncertainty $\delta$ is modelled to match the observational uncertainties. Therefore, we assigned a random value drawn from a Gaussian distribution centred at 0, with a standard deviation equal to the observational uncertainty specific to that observed star and epoch. This procedure ensures that the mock dataset retains the same measurement uncertainty characteristics as the observed data.

For each observed star and for each binary fraction $f$, we performed 10,000 iterations of the $v_{los}$ assignment. Finally, we computed the $\chi^2$ statistic for each mock star in each iteration.

\subsection{Comparing with observations}

Armed with this mock dataset, we now want to find the simulated distribution that best matches our observed distribution. In order to do that, we closely follow the approach of \cite{Arroyo_Polonio_2023}. We adopt the same likelihood function and use it to compare the cumulative distribution function (CDF) of $\chi^2$ calculated from the observations against those calculated from the mock datasets created assuming different binary fractions $f$. 
Therefore, conditioned on the observed data $D$ and the adopted model $M$, we defined the likelihood of a given binary fraction $f$ as
\begin{equation}\label{eq:likeli}
    P(f | D, M) \propto \left(\sum^{10000}_{i=1} \int \left|CDF_{obs} ( \chi^2 \leq X^2) - CDF_{sim}^{f,i} ( \chi^2 \leq X^2) \right| dX    \right)^{-1}.
\end{equation}
Here, the likelihood decreases as the difference between the observed cumulative distribution function (CDF) and the simulated CDFs increases, with the integral quantifying their distance and the sum averaging over 10,000 mock realisations at binary fraction $f$.
We compute the integral numerically by summing the differences in the observed and mock CDFs per observed $\chi^2$ value. 

To estimate the uncertainties associated with our fit results, we employed a bootstrap-like approach. We generated 1000 independent mock datasets for each tested $f$ and applied the identical fitting methodology used for the observed Leo T sample to each of these synthetic datasets.
We then identified the true $f$ of those mock datasets that yielded fit results statistically consistent with the results obtained from the observed data. The distribution of these true $f$ values was used to determine the 68\% highest posterior density (HPD) interval, which we report as the credible interval for our measurement.

\subsection{Binary distributions}
\label{sec:BinDists}

To quantify the velocity variations between epochs of observation that are caused by the orbital movement of a binary star, we performed a kinematic binary analysis. We operated under the assumption that a binary system can be treated as a two-point mass distribution governed solely by gravitational interaction, with radial velocity governed by the radial velocity equation, which is discussed in Appendix~\ref{sec:RVE}. 
This equation involves seven free parameters, four intrinsic to the binary system, and three extrinsic (which are discussed in Appendix~\ref{sec:RVE}). 
The three extrinsic parameters are the inclination ($i$), the argument of periapsis ($\omega$), and the true anomaly ($\nu$). 
The four intrinsic parameters are: $M$, the mass of the primary star; $q = \tfrac{m}{M}$, the mass ratio, where $m$ is the mass of the secondary star; $e$, the eccentricity of the orbit around the centre of mass, defined as $e = \sqrt{1-b^2/a^2}$ ($a$ being the semi-major axis and $b$ the semi-minor axis); and $P$, the period of the system. 
The intrinsic parameters set the physical properties of the binary system, while the extrinsic parameters determine the geometric projection that affects the observed radial velocity.
This section details the distributions used to simulate these four intrinsic binary parameters and explains our method for simulating $v_b$.

\subsubsection{Parameters distributions}
\label{sec:ParDist}

The primary challenge in this methodology lies in simulating a binary population with parameters that accurately represent what we are observing. Given that the study of binaries within these systems is still in its infancy, there is a lack of definitive guidance regarding the expected distribution of the parameters under discussion in this section. 

As a solution, we used the values of $M$ for the individual stars, as computed in Section~\ref{sec:spexxy_masses}, and combined them with the distributions outlined in \cite{Moe_2017}, which meticulously incorporated joint distributions of various orbital elements based on recent observations for the Milky Way. In our simulations, each star is therefore assigned its own fixed value of $M$, while the joint distribution of $P$, $q$, and $e$ is taken from \cite{Moe_2017}.
We refrain from delving into these distributions here; interested readers can refer to \cite{Moe_2017} for details.

\begin{center}
\begin{figure}[]
\resizebox{\hsize}{!}{\includegraphics[scale=1]{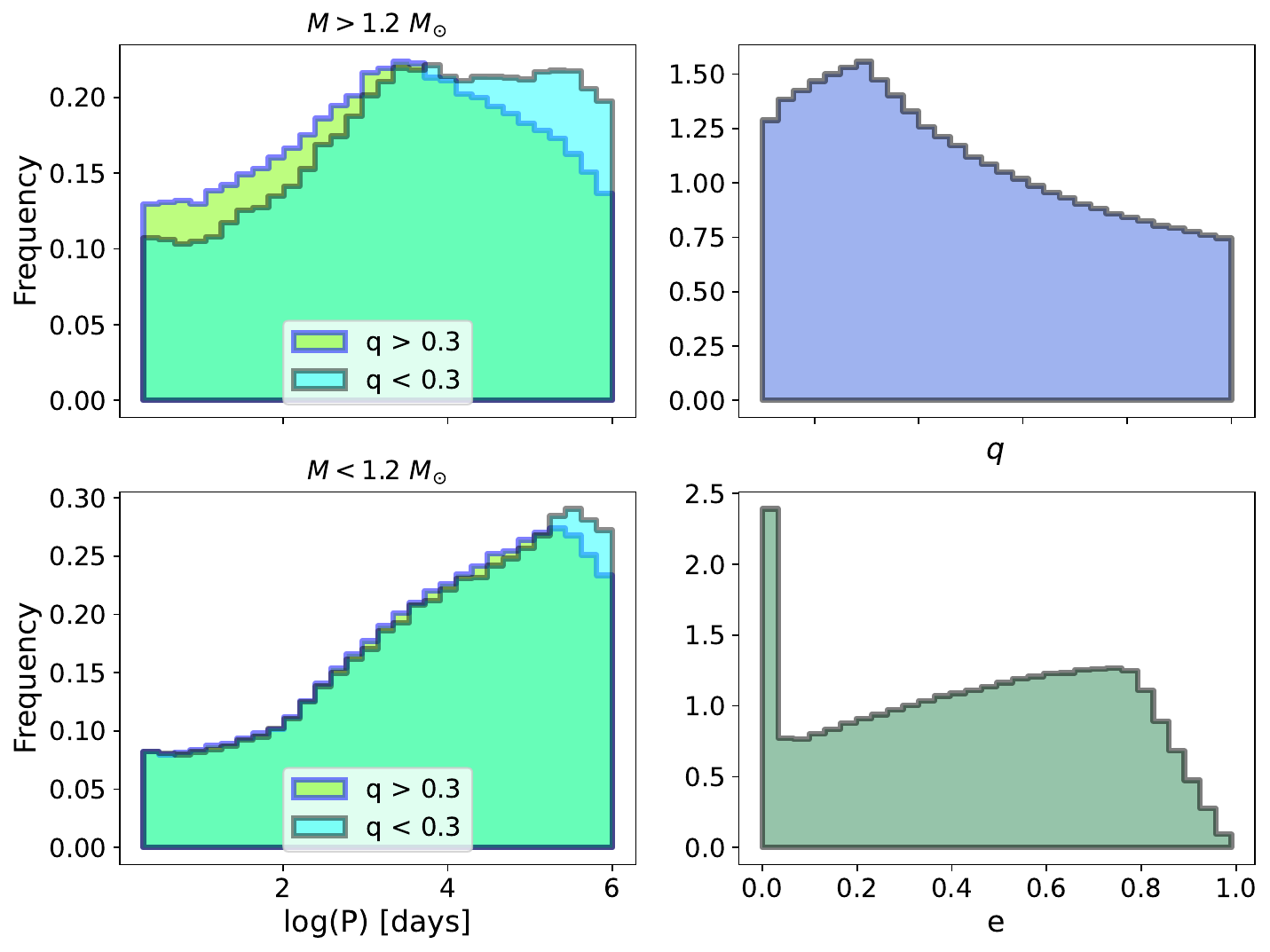}}
        \caption{Distributions of intrinsic parameters depicted in joint distributions showcasing various regimes for the period $P$ on the left. We show the distribution alterations with respect to the mass ratio $q$ and the primary mass $M$. The upper-left plot illustrates the distribution for $M > 1.2~M_{\odot}$, while the lower left pertains to $M < 1.2~M_{\odot}$, each highlighting regimes for $q > 0.3$ and $q<0.3$. 
        The top-right plot presents the distribution for the mass ratio q, while the bottom right illustrates the eccentricity distribution, notably demonstrating an abundance of circular orbits.}
        \label{fig:intdists}
\end{figure}
\end{center}

To implement these distributions, we used and customised the COSMIC binary population synthesis suite \citep{Breivik_2020} when necessary. Specifically, we used the multidimensional sampler, which is essentially a Python adaptation of the original IDL code from \cite{Moe_2017}. Note that multiple regimes per variable are common. For example, the variable $q$, which is sampled between $q=0.1$ and $q=1$ and is dependent on both $P$ and $M$, has distinct regimes, such as for $q < 0.3$ and $q \geq 0.3$, each delineating specific conditions within the distribution. Another regime is that of $q>0.95$, since it is explicitly considered the excess probability of equal-mass binaries. We disregard this excess probability, as it is unlikely to persist into later evolutionary stages. Many of such binaries are expected to have radii that surpass the Roche-lobe size, making a large fraction of twin binaries in old dwarf systems improbable. This is the only modification that we implemented regarding the distribution related to $q$, which is shown in Figure~\ref{fig:intdists} (top right corner). 

Turning to the parameter $P$, its dependence on both $M$ and $q$ introduces multiple regimes within both parameters. We adapted the distribution of $P$ with the nuance of constraining the range for which the parameter is sampled.
To start with, we restricted the period distribution to $P \leq 10^6$ days. This corresponds to orbital separations of roughly 200~AU, marking the transition to the wide-binary regime. At such separations the velocity amplitudes drop below $\sim$1~km\,s$^{-1}$, rendering these systems undetectable with our method and irrelevant for velocity-dispersion inflation. Wide binaries are also expected to be dynamically fragile, and their survival rate in dwarf galaxies remains poorly constrained. Since our analysis focuses on binaries capable of affecting the measured velocity dispersion and given the lack of data to model the wide-binary regime reliably, we excluded systems with $P > 10^6$ days.

In order to estimate the close binary fraction, we limited the semi-major axis, $a$, to $a \leq 10$ au, artificially limiting the period distribution to $P \lesssim 10^4$ days.
Finally, for all of the above, the minimum period allowed was $P=2$ days, which is the typical length of an OB in our observation, and also reduces the number of circularised binaries, while limiting the number of extremely close binaries that would be disrupted due to Roche-lobe overflow. The distributions of $P$ are shown in Figure~\ref{fig:intdists}, in the left corner. Two regimes are shown, for $M > 1.2~ M_\odot$, on the top, and for $M < 1.2~ M_\odot$, on the bottom.

The parameter $e$ depends on $M$ and, more notably, on $P$. When $P\leq2$ days, it is assumed that all binaries are circularised. For $P>2$ days there is an upper limit on $e$ which is
$e_{max} = 1 - \left(\frac{P}{2~\mathrm{days}}\right)^{-2/3} $, which guarantees that the binaries have Roche-lobe fill-factors $\leq 70\%$ at periapsis. The distribution of $e$ is shown in Figure~\ref{fig:intdists}, in the bottom right corner. 

\section{Results and discussion}
\label{sec:results}

In this section, we present the results derived from the methodology described in Section~\ref{sec:FM}. Our analysis is divided into several steps. We begin by estimating the total binary fraction of Leo T by comparing our observations with a simulated binary population covering periods P $\leq 10^6$ days, and comparing these results with those found in the literature. 

However, the magnitude of our measurement uncertainties restricts our sensitivity to binaries with low radial velocity amplitudes. This observational limitation introduces degeneracies in our modelling, particularly for long-period systems where orbital velocities are naturally lower. Nevertheless, it is important to note that even stars with larger measurement errors contribute meaningfully to this forward modelling analysis. By effectively constraining the high-amplitude tail of the velocity distribution, we can statistically infer the underlying binary fraction based on the assumed orbital distributions.

To mitigate these degeneracies and focus on the regime where our sensitivity is highest, we adopt a two-pronged approach. First, we shift our attention to analysing the close binary fraction of Leo T. This involved limiting the distribution of the semi-major axis to $a < 10~\mathrm{au}$. Given the distinct stellar populations within Leo T, we performed the same analyses for each independently to explore whether they yielded consistent results for $f$.

Subsequently, we limit the binary population of our mock datasets by only allowing binaries with a semi-amplitude of the radial velocity $\geq 10~\mathrm{km\ s^{-1}}$, which is the typical value for our velocity uncertainty measurements. As such, binaries whose radial velocity semi-amplitudes are of this order would inflate our measurement of the system’s velocity dispersion.
Finally, we discuss these results in light of their potential impact on our velocity dispersion estimations. 

\begin{center}
\begin{figure}[]
\resizebox{\hsize}{!}{\includegraphics[scale=1]{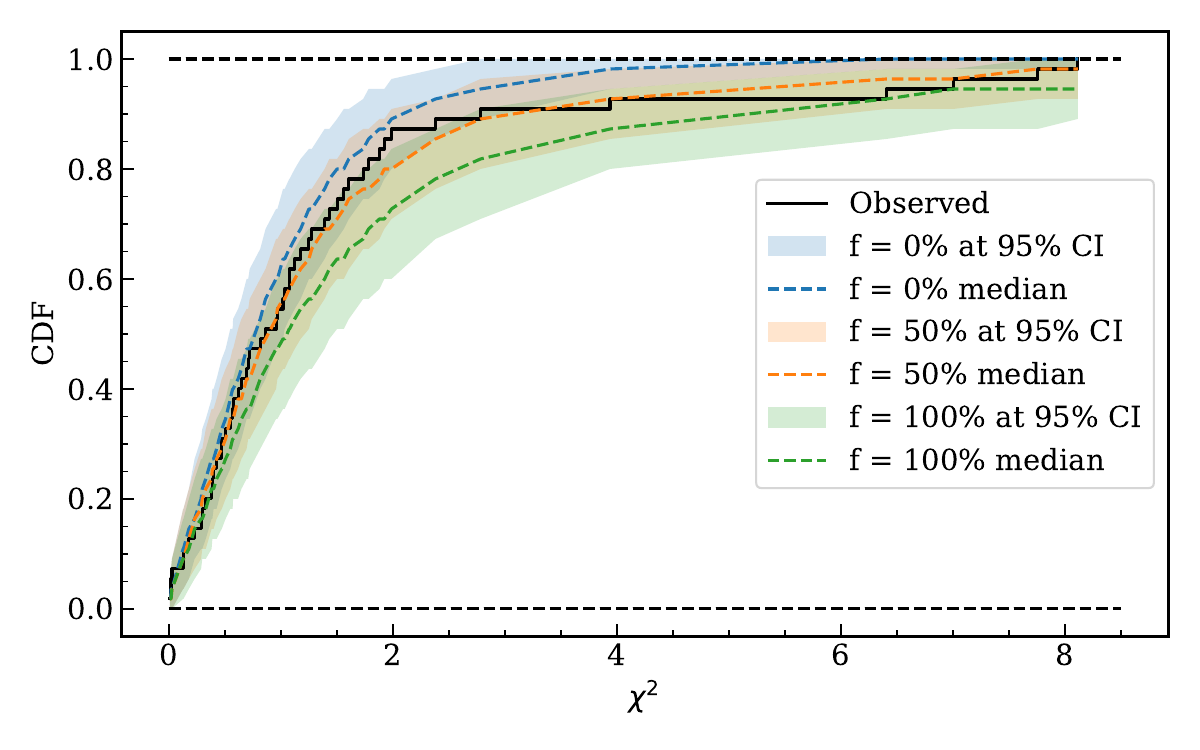}}
        \caption{Cumulative distribution functions of the $\chi^2$ statistic for Leo~T. The solid black line shows the observed CDF. Shaded regions indicate the 95\% confidence intervals of the mock CDFs for $f_{\mathrm{mock}} = 0$\%, 50\%, and 100\%, derived from 10,000 realisations. Dashed curves show the corresponding median mock CDFs.}
        \label{fig:CDF}
\end{figure}
\end{center}

\subsection{Binary fraction of Leo T}\label{sec:DiscFrac}

We started by comparing the full observed sample with the complete mock binary population, allowing for periods $P \leq 10^6$ days, which translates to semi-major axes reaching as high as $370~\mathrm{au}$ in our configuration.
Using Equation~\ref{eq:likeli}, we obtained a maximum probability for $30^{+22}_{-5}$ binaries in the sample, which corresponds to a binary fraction of $55^{+40}_{-9} \%$. As anticipated, the upper bound is not well constrained, mainly due to degeneracies. This arises from the challenge of distinguishing non-binaries from binaries with long orbital periods, where the velocity amplitude induced by orbital motion is too small to detect. Orbital inclination also plays a role in this, as a binary system with a face-on orbit or one close to it becomes undetectable in velocity variation studies. We try to mitigate these effects further below.
For visual reference, Figure~\ref{fig:CDF} compares the observed CDF of the $\chi^2$ statistic (solid black line) with mock CDFs generated for $f_{\mathrm{mock}} = 0$\%, 50\%, and 100\%. Each shaded band represents the 95\% confidence interval, computed from 10,000 realisations by taking the 2.5th and 97.5th percentiles of the mock CDFs, while the dashed curves trace their medians. The separation between the bands reflects the sensitivity provided by Leo~T’s multi-year coverage, while the observed CDF, though noisier, falls closest to the $f_{\mathrm{mock}} = 50$\% case, consistent with our best fit binary fraction of $f \simeq 0.55$.

The binary fraction of $55^{+40}_{-9} \%$ that we find for Leo T can be compared with other values in the literature. It is commonly accepted that more than $50\%$ of stars exist within multiple systems \citep[e.g.][]{Duchene_2013}, and the value we find here for Leo T agrees with this. This value mirrors the typical occurrence observed in luminous dwarf spheroidals as well \citep{Minor_2013}.

\begin{table}[ht]
\footnotesize
\centering
\caption{List of binary fractions with $1\sigma$ uncertainty from the literature with the new estimation for Leo T at the end.}
\begin{tabular}{lcccccrr}
    \hline\hline
    \multicolumn{1}{c}{Galaxy} & \multicolumn{1}{c}{f} & \multicolumn{1}{c}{$1 \sigma$} & \multicolumn{1}{c}{$M_V$} & \multicolumn{1}{c}{$[Fe/H]$} \\ 
    \hline
    Sculptor      & $0.59^1$                 & [0.43, 0.83]             & $-11.1  \pm 0.5~^5$ & $-1.68~^5$ \\
                  & $0.58^2$                & [0.41, 0.73]              &  \\
                  & $0.55^3$                  & [0.37, 0.72]             &  \\
    \hline
    Draco         & $0.50^2$                  & [0.46, 0.55]            &  $-8.8 \pm 0.3~^5$ & $-1.93~^5$  \\
                  & $0.49^3$                  & [0.28, 0.72]              &  \\
    \hline
    Ursa Minor    & $0.78^2$                  & [0.69, 0.86]            &  $-8.8 \pm 0.5~^5$ & $-2.13~^5$  \\
                  & $0.67^3$                  & [0.38, 0.88]             &  \\
    \hline
    Carina        & $0.14^1$                 & [0.09, 0.42]             & $-9.1 \pm 0.5~^5$ & $-1.72~^5$  \\
                  & $0.20^2$                & [0.07, 0.29]              &  \\
                  & $0.25^3$                  & [0.07, 0.56]              &  \\   
    \hline
    Sextans       & $0.69^1$                  & [0.46, 0.89]              & $-9.3 \pm 0.5~^5$ & $-1.93~^5$  \\
                  & $0.71^2$                & [0.57, 0.73]              &  \\
                  & $0.55^3$                  & [0.25, 0.81]              &  \\   
    \hline
    Fornax        & $0.44^1$                 & [0.32, 0.70]              & $-13.4 \pm 0.3~^5$ & $-0.99~^5$  \\
                  & $0.87^2$                & [0.78, 0.99]              &  \\
                  & $0.66^3$                  & [0.34, 0.90]              &  \\   
    \hline
    Leo II        & $0.36^2$                  & [0.28, 0.43]              & $-9.8 \pm 0.3~^5$ & $-1.62~^5$  \\
                  & $0.40^3$                  & [0.19, 0.62]              &  \\
    \hline
    Reticulum II  & > $0.50^{4}$            & -              &  $-2.7 \pm 0.1~^6$ & $-2.65~^7$  \\
    \hline
    Leo T       & 0.55                  & [0.46, 0.95]              &  $-8.0 \pm 0.5~^5$ & $-1.6$ \\
    \hline
    \end{tabular}
\label{tab:BinFrac}
\tablebib{$^1$\citet{Minor_2013}; $^2$\citet{Spencer_2018};
  $^3$\citet{Arroyo_Polonio_2023};
  $^4$\citet{Minor_2019}; $^5$\citet{McConnachie_2012}; $^6$\citet{Koposov_2015}; $^7$\citet{Simon_2015}.}

\end{table}

In Table~\ref{tab:BinFrac}, multiple estimated binary fractions for various galaxies are presented. We highlight that the results shown here for \cite{Arroyo_Polonio_2023} are the ones obtained by using the new methodology presented in their work, incorporating Roche-lobe overflow in the estimation process.

Most of the examples show binary fractions that typically exceed 50\%. For example, the binary fraction of Sculptor has consistently been estimated around $\sim 60\%$, as supported by the findings of \cite{Minor_2013,Spencer_2018, Arroyo_Polonio_2023}.
Even the faintest galaxy in this sample, Reticulum II, exhibits a binary fraction higher than 50\% at the 90\% confidence level (assuming a mean orbital period of 30 years or longer). However, the Carina dwarf appears to be an exception, showing a noticeable deficiency in binary stars compared to the others in the analysis.

\subsection{Close binary fraction}

The definition of close binaries lacks a universally fixed criterion. We can typically see that this designation refers to binary systems with separations of $\leq 10$ au \citep[e.g.][]{Kratter_2011, Offner_2023}, and this is the definition that we adopt here. 

Importantly, the association between the close binary fraction and metallicity has been a subject of debate. Some observations suggest that there is no direct link between these factors \citep[e.g.][]{Moe_2013}, while others indicate a positive correlation between the close binary fraction and metallicity \citep[e.g.][]{Hettinger_2015}. In contrast, there are findings suggesting that the close binary fraction is strongly anti-correlated with metallicity \citep[e.g.][]{Raghavan_2010, Badenes_2018, Moe_2019}. Demonstrating this anti-correlation presents challenges due to the sensitivity of most binary detection methods, which often vary with metallicity. However, a reanalysis of binary populations in five surveys of \cite{Moe_2019} showed that after correcting for observational biases, the five surveys showed consistent indications of a metallicity-dependent close binary fraction, with the latter decreasing from $53 \pm 12\%$ at $\mathrm{[Fe/H]} = -3$ to $10 \pm 3\%$ at $\mathrm{[Fe/H]} = +0.5$. 

We can further analyse our data by imposing a restriction on the allowable mock binaries, limiting their orbital semi-major axis to $\leq 10$ au. This constraint consequently restricts the probed period to $P \lesssim 10^4$ days.
The study by \cite{Moe_2019} reported close binary fractions of $53 \pm 12\%$ at [Fe/H] = -3 and $40 \pm 6\%$ at [Fe/H] = -1. Here, we obtain a maximum probability for $16^{+19}_{-5}$ binaries in the sample, which corresponds to a close binary fraction of $30^{+34}_{-9} \%$ at [Fe/H] $\sim -1.6$, a value consistent with the findings of \cite{Moe_2019}. 

We note that the measurements in \cite{Moe_2019} were focused on Solar-type stars, which differed from our sample composition. Furthermore, our sample comprises a mix of young and old stars, which may influence our results, as discussed in Section~\ref{sec:yo}.

\subsubsection{Young stars versus old stars}\label{sec:yo}

One of the most intriguing aspects of Leo T is its recent star formation, evident in the current presence of two distinct stellar populations, a feature notable in our sample. 
Differences in the kinematics of these populations have also been discussed in detail in \citetalias{Vaz_2023}.

To investigate potential differences in binary fractions between these populations, we repeat the analysis, this time dividing the observed sample into two groups: 20 stars classified as young and the remaining 35 as old. 
We obtain a close binary fraction of $35^{+40}_{-6} \%$ for the young population, corresponding to $7^{+8}_{-2}$ binaries in the sample.
For the old population, we get a close binary fraction of $15^{+43}_{-15}\%$, corresponding to $5^{+15}_{-5}$ binaries in the sample.

The higher close binary fraction we observed in the younger stellar population of Leo T compared to the older population aligns with expectations. Stellar multiplicity is known to increase with primary mass \citep{Offner_2023}, and the younger stars in Leo T are typically more massive than the older ones, naturally explaining this difference.

Furthermore, over time, binary systems can be disrupted, potentially lowering the binary fraction among older populations. Stellar evolution, particularly the expansion in later stages, can lead to Roche-lobe overflow in close binaries, resulting in mass transfer that may destabilise the system. 
Roche-lobe overflow is not the only factor that contributes to binary disruption. Research has indicated that the evolution of binaries within UFDs involves a complex interplay of tidal effects induced by the dark matter halo and other collisional interactions resulting from encounters between binary - binary systems or binary - single stars \citep{Livernois_2023}. These effects notably amplify the disruption of wide binaries and leave discernible traces on the semi-major axis distribution of the surviving binaries, often causing a significant migration of binaries towards closer orbits. However, the precise distributions of the period and semi-major axes remain elusive because of these intricate evolutionary mechanisms. Consequently, more in-depth studies are imperative to refine and enhance our understanding and measurements of these parameters. In that sense, Leo T indeed emerges as an intriguing case for exploring how binaries are disrupted in such systems. Its multiple stellar populations offer an exceptional opportunity to quantify the process of binary disruption. More extensive and higher-quality data could shed more light on the intricacies of binary evolution within Leo T.

Another intriguing aspect of these findings is the observed lower velocity dispersion of $\sigma_{v} = 2.31^{+2.68}_{-1.65}\ \mathrm{km}\,\mathrm{s}^{-1}$ for the young population in \citetalias{Vaz_2023}. To discern whether this value might be influenced by the inclusion of binaries in our sample, a meticulous analysis is necessary. We discuss this in Section~\ref{sec:veldisp}.

\subsection{Impact on the inferred velocity dispersion}\label{sec:veldisp}

\subsubsection{Limiting the semi-amplitude}

One critical aspect of measuring the binary fraction is understanding its impact on the measured velocity dispersion. It is important to note that only binaries with substantial velocity amplitudes would significantly affect our velocity dispersion measurements, considering the relatively high uncertainty in our velocity measurements. Hence, in our analysis, we constrain even further the maximum allowable period for our mock sample.
We repeated the comparison between the mock sample and the observed one but excluded semi-amplitudes $\leq~10~\mathrm{km\ s^{-1}}$, as this criterion provides the most reliable means of identifying binaries within our sample. Given that our velocity uncertainties typically hover around $15~\mathrm{km\ s^{-1}}$ and the available epochs are limited to $\leq 5$, our dataset lacks the volume and precision necessary to detect subtle velocity variations between epochs. 
Our findings yield a maximum probability for $11^{+14}_{-2}$ binaries in the sample, corresponding to a binary fraction of $20^{+26}_{-2} \%$.

We also performed this analysis for the old and young populations separately. Again, the maximum probability for the young population indicates a higher binary fraction than that for the old population. For the old population, the maximum probability corresponds to a binary fraction of $10^{+19}_{-10}\%$, corresponding to $3^{+7}_{-3}$ binaries in the sample.

In contrast, the maximum probability for the young population corresponds to a binary fraction of $30^{+30}_{-15} \%$, corresponding to $6^{+6}_{-3}$ binaries in the sample. 
Again, we find a notable difference in the binary fraction between the old and young populations, in agreement with the results for the close binary fraction.

\subsubsection{Velocity dispersion calculation}

To conduct an analysis on the impact of binaries on velocity dispersion measurements, we selected the epoch with the highest S/N, which also provided reliable measurements for all 55 stars in our sample. Using the same methodology as in \citetalias{Vaz_2023}, we performed a kinematic analysis aimed at estimating the velocity dispersion from these 55 stars.
Subsequently, we repeated the analysis after excluding 11 stars from the sample, as suggested by our best-fit binary fraction. Specifically, we removed the stars with the largest $\chi^2$ values, since these are the most likely to be binaries in our framework. While not all of these stars have p-values that would independently identify them as variable, the majority show significant evidence of variability, and excluding them provides a consistency check on the robustness of the binary fraction estimate.
The corner plots that display these results are shown in Appendix~\ref{app:corner}.

For the whole sample, we get a velocity dispersion of $\sigma_{v} = 6.57^{+2.71}_{-2.44}\ \mathrm{km}\,\mathrm{s}^{-1}$.
By repeating the calculations but removing the 11 stars most likely to be binaries from the sample, we obtained a velocity dispersion of $\sigma_{v} = 7.21^{+2.72}_{-2.45}\ \mathrm{km}\,\mathrm{s}^{-1}$, very similar to what was obtained in \citetalias{Vaz_2023}.

It is intriguing that our correction for binaries does not significantly alter our results, especially in comparison to \citetalias{Vaz_2023}. This suggests that the results of \citetalias{Vaz_2023}, where all epochs were combined, already accounted for binary contamination. In our current study we analyse individual epochs, but even if there is no clear velocity offset between them, the combined spectrum of a binary can differ from that of any single epoch. Attempting to fit such combined spectra can in some cases lead to failed fits \citep{Badry_2017}, which may explain why certain stars included in our present analysis were not part of the final sample in \citetalias{Vaz_2023}. For instance, the star excluded in \citetalias{Vaz_2023} due to velocity-estimation errors is one of the most probable binaries in our current sample, ranking as the star with the second highest $\chi^2$ value. In total, four stars that are now identified as likely binary candidates were excluded from \citetalias{Vaz_2023} because their fits did not converge.

Beyond these special cases, the dominant effect is that co-adding spectra with different velocities broadens the absorption lines, and the full-spectrum fit then centres the model on the mean velocity of the input epochs. 
To quantify this effect, we constructed a suite of mock spectra (detailed in Appendix~\ref{apx:test_spexxy}) to verify that co-adding multi-epoch observations of binary stars yields line-of-sight velocities equal to the average of the combined epochs. We find that this holds true for the vast majority of cases. Only in extreme scenarios, where double-peaked profiles arise from very large velocity separations ($\sim$150--200 km\,s$^{-1}$), does the fit occasionally converge on a single component rather than the mean. While such separations would suggest the presence of compact object companions, we find no evidence of these extreme systems in our observed sample, nor are they expected to be present. Consequently, we argue that the co-added spectra effectively provide period-averaged stellar velocities for our sample, mitigating the impact of binaries on the velocity dispersion and explaining the consistency between our results and those of \citetalias{Vaz_2023}.

\subsubsection{Velocity dispersion calculation: Young stars versus old stars}

In a similar manner as above, we conducted identical analyses for both the young and old samples. Initially, we computed the velocity dispersion for a subset of 20 and 35 stars, representing the young and old samples, respectively. Subsequently, we excluded the most probable binary stars from each subset, leading to reduced samples comprising 14 and 32 stars for the young and old groups, respectively.
The corner plots that display these results are shown in Appendix~\ref{app:corner}.

We obtained velocity dispersions of $\sigma_{v_{young}} = 6.51^{+4.45}_{-3.60}\ \mathrm{km}\,\mathrm{s}^{-1}$ and $\sigma_{v_{old}} = 5.32^{+4.23}_{-3.42}\ \mathrm{km}\,\mathrm{s}^{-1}$, which are in line with what was obtained for the full sample, but not with what was obtained in \citetalias{Vaz_2023} (see Table~\ref{tab:LeoTP}). 
We note that the high uncertainty of these measurements is expected and results from the smaller samples. 

If we now remove the stars identified as the most probable binaries from each sample, we get $\sigma_{v_{young}} = 4.07^{+4.48}_{-2.81}\ \mathrm{km}\,\mathrm{s}^{-1}$ and $\sigma_{v_{old}} = 4.85^{+4.63}_{-3.32}\ \mathrm{km}\,\mathrm{s}^{-1}$.
Here, we see a deflation of the measured velocity dispersion, more noticeable in the young sample, as expected from the higher binary fraction in that group. However, the velocity dispersion of the young sample still seems to be inflated compared to the results of \citetalias{Vaz_2023}, but the high uncertainty of our measurement here means that both measurements are consistent. 
This finding supports our conclusion that the velocity dispersion measurements in \citetalias{Vaz_2023} were largely unaffected by binary orbital motion. Furthermore, our results reinforce the efficacy of co-adding multi-epoch spectra as a robust method for mitigating the impact of binaries on line-of-sight velocity measurements.

\begin{figure}[t]
  \centering
  \includegraphics[width=\linewidth]{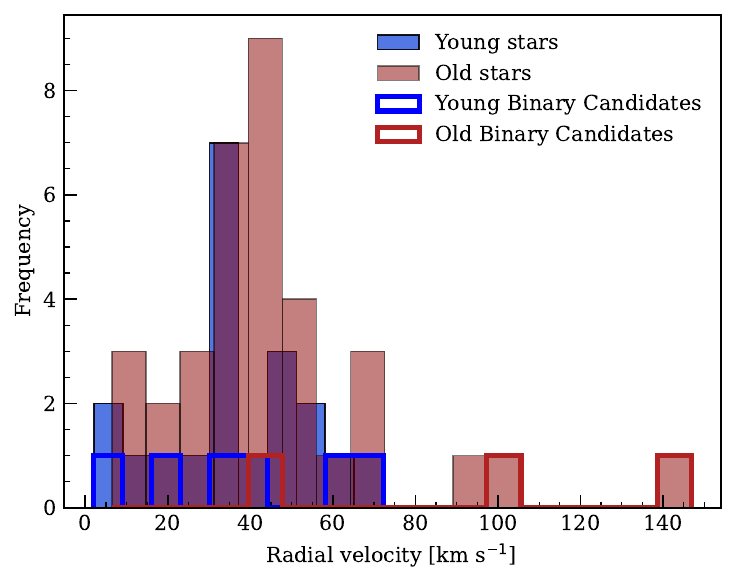}
  \caption{Radial velocity distributions of the young (blue) and old (red) stellar populations in Leo~T. Filled histograms show the full samples, with five bins for the young population and ten bins for the old population. Outliers identified by their high $\chi^2$ values (6 for the young stars and 3 for the old stars) are indicated by the step histograms with bold outlines. The mean velocities of the two populations differ by more than $1\sigma$, consistent with the presence of two kinematically distinct groups.}
  \label{fig:vel_hist}
\end{figure}

Another result that we obtain from these analyses is that the older population also appears to have a lower velocity dispersion than previously estimated. We also note that the mean velocity of each population differs, with the younger population having a best fit of 
$v_{\mathrm{young}} = 34.05^{+3.15}_{-3.07}\ \mathrm{km\,s}^{-1}$, while for the older population we measured $v_{\mathrm{old}} = 43.04^{+2.33}_{-2.78}\ \mathrm{km\,s}^{-1}$. These values are inconsistent at more than $1\sigma$, suggesting that the two populations are indeed centred at different systemic velocities. As a result, the velocity dispersion of the combined population reported above is likely inflated by the mixing of stars from two kinematically distinct groups.  
To illustrate this, Figure~\ref{fig:vel_hist} presents the velocity histograms of the young and old populations. The high-$\chi^2$ outliers excluded from the analysis are also shown, demonstrating that the offset in systemic velocities between the two groups is not driven by a few extreme cases but is present in the overall distributions.

\section{Conclusions}
\label{sec:conclusions}

In the quest for precision in measuring the velocity dispersion of faint galaxies, the influence of binary stars emerges as a key challenge, potentially impacting observed velocity dispersions. Here we present the first ever measurement of the binary fraction for Leo T using multi-epoch data from the MUSE-Faint survey.
We obtained a best-fit binary fraction of $55^{+40}_{-9} \%$, which aligns with estimates within the Milky Way, where more than 50\% of stars are estimated to exist in multiple systems. This consistency extends to other dwarf galaxies, whose binary fractions often exceed 50\%, as seen in prior research such as \cite{Minor_2013}.
However, our measurements carry a substantial degree of uncertainty, attributed to the limited number of observed stars, higher uncertainties in velocity measurements, and a restricted number of useful observed epochs per star — typical challenges for these faint systems. Compounding this uncertainty is the inherent degeneracy, particularly at higher period values, where velocity variations between epochs become negligible and evade detection through our methods. Furthermore, the lack of extensive studies on the period distribution and other intrinsic parameters in these systems introduced further limitations to our study.

To mitigate these constraints, we focused on restricting the period distribution to a maximum value; reducing degeneracy, which occurs primarily at higher periods; and aligning the possible period distributions, where agreement tends to converge for lower periods. Thus, we estimated the close binary fraction of Leo T by limiting the maximum allowed semi-major axis to $a < 10~\mathrm{au}$, for which we obtained the maximum likelihood for a close binary fraction of $30^{+34}_{-9} \%$. Our findings are consistent with the literature discussing the anti-correlation between close binary fractions and metallicity (for solar-type stars), though our measurement carries a high uncertainty. In particular, \cite{Moe_2019} reports close binary fractions of $53 \pm 12\%$ at $\mathrm{[Fe/H] = -3}$ and $40 \pm 6\%$ at $\mathrm{[Fe/H] = -1}$, while here we measured $30 \pm 24\%$ at $\mathrm{[Fe/H]} = -1.6$.

Given the presence of two distinct stellar populations in Leo T, we extended our analysis to determine the close binary fraction for each population separately.
Our measurements yielded a close binary fraction of $35^{+40}_{-6}$ for the younger population and $15^{+43}_{-15}\%$ for the older population.
This significant difference is primarily attributed to the well-established correlation between stellar multiplicity and primary mass \citep{Offner_2023}, with the younger population in Leo T typically comprising more massive stars. Furthermore, evolutionary processes and the unique environment of UFDs likely contribute to the lower binary fraction in the older population. Close binaries are susceptible to disruption in later stellar evolutionary stages due to Roche-lobe overflow, where an expanding star transfers mass to its companion, potentially destabilising the system. 
Additionally, the high dark matter density within UFDs suggests dominance by tidal effects originating from the dark matter halo. This, coupled with other collisional interactions, intensifies the disruption of wider binaries and leaves observable imprints on the semi-major axis distribution of surviving binaries \citep{Livernois_2023}, which often leads to a substantial migration of binaries towards closer orbits. These combined effects likely contribute to the observed lower close binary fraction in the older stellar population.

Furthermore, we analysed whether the velocity dispersion measurements reported in \citetalias{Vaz_2023} might have been inflated by the presence of unresolved binary stars in their sample. Given the limitations of our dataset in identifying binaries with velocity variations not exceeding approximately $15~\mathrm{km\ s^{-1}}$ between observed epochs, we restricted the allowed semi-amplitude distribution in our mock data to a minimum of $10~\mathrm{km\ s^{-1}}$. This particular restriction allowed us to identify the most plausible binary candidates in our sample, and we were then able to exclude them from our velocity dispersion analysis.
Surprisingly, our correction for binaries did not exhibit a discernible difference in velocity dispersion compared to the original methodology. However, when examining the young and old populations separately, we observed an inflation in the measured velocity dispersion specifically within the young population, aligning with the higher estimated binary fraction in this subgroup.

These findings suggest that our measurements in \citetalias{Vaz_2023} might have already accounted for the inflationary effect of binary stars. The analysis in that particular study involved combining multi-epoch data into a single data cube, resulting in certain stellar spectra failing the quality control check due to challenges in fitting the stellar parameters that did pass the quality criteria when analysed for each epoch separately. Moreover, the velocity measurements for the multi-epoch combined data cube are expected to be an average of the velocity of the epochs being combined. These factors might have inadvertently addressed the binary star effects on the velocity dispersion measurements.

In conclusion, our findings underscore the intricate dynamics of binaries in these faint systems, demonstrating their complex evolutionary phases and susceptibility to diverse interactions with the environment. There is a crucial need for extensive and higher-quality data of multiple dwarf galaxies, as it helps us have a deeper understanding of the period and other parameter distributions of binaries in these dark matter-dominated systems. It is only with such data that we can aim to comprehend how binary systems evolve in dark matter-dominated environments and their influence on the accurate estimation of dynamical masses. Future research with larger datasets and improved observational capabilities will be essential to further unravelling these complexities.

\begin{acknowledgements}

DV and JB acknowledge support by Fundação para a Ciência e a Tecnologia (FCT) through the research grants
UID/FIS/04434/2019, UIDB/04434/2020, UIDP/04434/2020. DV acknowledges support from the Fundação para
a Ciência e a Tecnologia (FCT) through the Fellowship 2022.13277.BD.
JB acknowledges financial support from the Fundação para a Ciência e a
Tecnologia (FCT) through national funds PTDC/FIS-AST/4862/2020 and
work contract 2020.03379.CEECIND. SKA gratefully acknowledges funding from UKRI through a Future Leaders Fellowship (grants MR/T022868/1, MR/Y034147/1). SS acknowledges funding from the European Union under the grant ERC-2022-AdG, ‘StarDance: the non-canonical evolution of stars in clusters’, grant agreement 101093572, PI: E. Pancino. PACC acknowledges financial support from
Fundação para a Ciência e Tecnologia (FCT) through grant 2022.11477.BD.
Based on observations made with ESO Telescopes at the
La Silla Paranal Observatory under programme IDs 0100.D-0807, 0101.D-0300,
0102.D-0372 and 0103.D-0705. This research has made use of Astropy \citep{astropy, astropy2018}, corner.py \citep{FM2013}, matplotlib \citep{Hunter:2007}, NASA’s Astrophysics Data System Bibliographic Services, NumPy \citep{harris2020array}, SciPy \citep{2020SciPy-NMeth}.

\end{acknowledgements}

%
%
\bibliographystyle{aa}
\bibliography{biblio}

\begin{appendix}

\section{The radial velocity equation}
\label{sec:RVE}

We start by expressing the orbital radial velocity equation, describing the radial velocity component of a star orbiting another body as observed by an observer, as
\begin{equation}\label{eq:RVE}
V_r = \frac{q \sin i}{\sqrt{1-e^2}} \left[ \frac{2\pi G M}{P (1+q)^2} \right]^{1/3} \left(\cos (\omega + \nu) + e \cos \omega \right).
\end{equation}
The equation has seven parameters: $M$ is the mass of the primary star; $q = \frac{m}{M}$ is the ratio of the mass of the secondary star, $m$, and that of the primary star; $e$ is the eccentricity of the orbit around the centre of mass, which is defined as $e = \sqrt{1-b^2/a^2}$, where $a$ is the semi-major axis and $b$ is the semi-minor axis; $P$ is the period of the system; $i$ is the inclination, the angle that orbital plane makes with the plane of the sky;
$\omega$ is the argument of periapsis, which is the angle corresponding to the point of the orbit that is closest to the centre of mass, as it is the angle between the line connecting the ascending node to the focus and the focus to the periapsis; $\nu$ is the true anomaly, which is the angle between the line connecting the periapsis to the focus and the focus to the star. The latter stands as the sole parameter that dynamically shifts with time, serving as an indicator of the position of the star along its orbit.

Of these seven parameters, four are intrinsic ($q, e, M$ and $P$) and three extrinsic ($i, \omega$ and $\nu$). We discuss the intrinsic parameters in detail in Section~\ref{sec:ParDist}.
The three extrinsic parameters are straightforward. We randomly drew them as follows for each star considered binary in the mock sample:

1) The angle corresponding to the inclination ($i$), from a distribution $f(i) = \sin i$, that ranges from $i=0$ rad, corresponding to a face-on orbit, to $i = \frac{\pi}{2}$ rad.

2) The argument of periapsis ($\omega$), from a uniform distribution that ranges from $0$ to $2\pi$ rad. 

\begin{center}
\begin{figure}[]
\resizebox{\hsize}{!}{\includegraphics[scale=1]{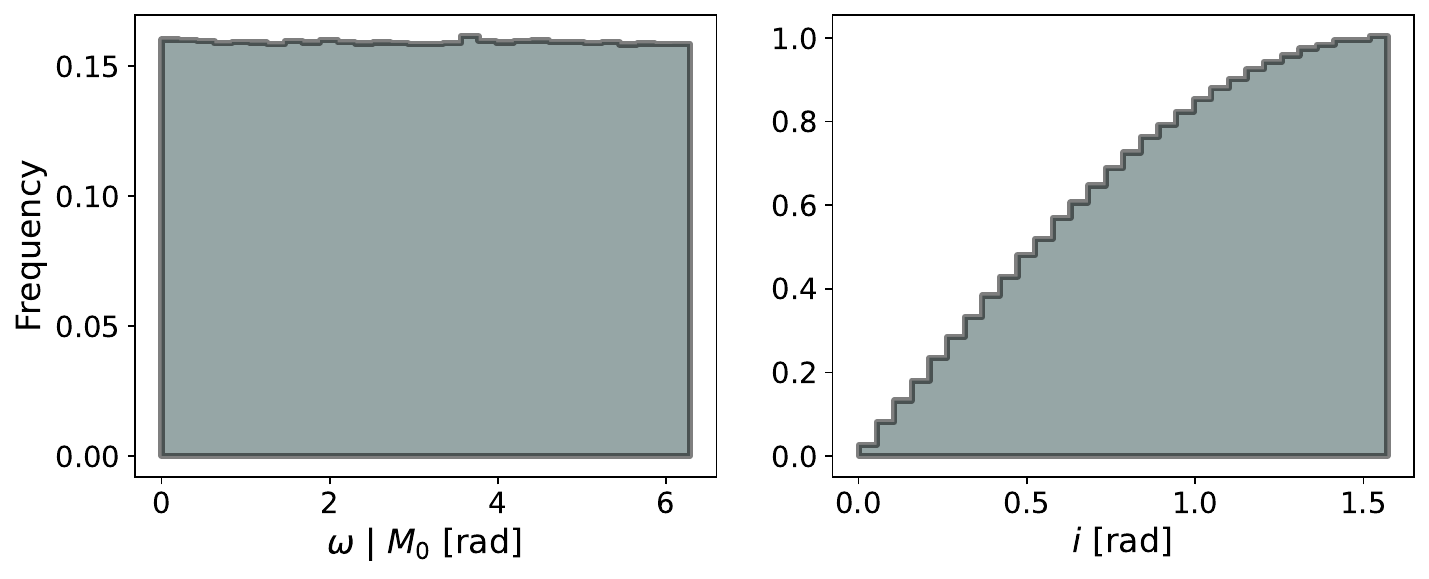}}
        \caption{Distributions of the extrinsic parameters. On the left, the distribution of $\omega$ is shown, which is the same distribution of $M_0$. On the right it is displayed the distribution of the inclination $i$.} 
        \label{fig:extdists}
\end{figure}
\end{center}

3) The true anomaly ($\nu$), for which we need to take into account the variation resulting from the time of each observation. We sample this parameter by making use of an auxiliary parameter called mean anomaly ($l$) which represents the fraction of an elliptical orbit's period that has elapsed after a given time, $t$, which we consider to be the time difference between epochs of observation. 
We can relate $l$ to $t$ by writing $l = l_0 + nt$, where $l_0$ is the initial mean anomaly and $n$ is the mean angular motion and is given by $n = \frac{2\pi}{P}$. Since $l_0$ is random, we can sample it from a uniform distribution ranging from $0$ to $2\pi$ rad. We can now compute the eccentric anomaly, $E$, which is related to $l$ through the expression $l = E - e\cdot \sin E$, where $e$ is the eccentricity. Finally, we can compute $\nu$ using the expression $\cos \nu = \frac{\cos (E) - e}{1-e\cdot \cos E}$. 
The distributions of the extrinsic parameters are illustrated in Figure~\ref{fig:extdists}. As the distributions of $\omega$ and $l_0$ are the same, we only show one distribution to represent both. 

\section{Uncertainties estimations}\label{app:violin}

A crucial aspect of our analysis concerns the uncertainties associated with our radial velocity measurements, as these fundamentally constrain the parameter space of detectable binaries. Previous studies have established that MUSE achieves a systematic accuracy limit of $\approx 1$~km\,s$^{-1}$ in the analysis of stellar spectra \citep{kamann16}. However, because the individual stars observed in Leo T are extremely faint, our derived uncertainties are significantly larger than this instrumental floor. Consequently, the systematic error is negligible compared to the statistical measurement errors, and we focus our assessment on the latter.

\begin{figure}[!h]
\centering
\includegraphics[width=\linewidth]{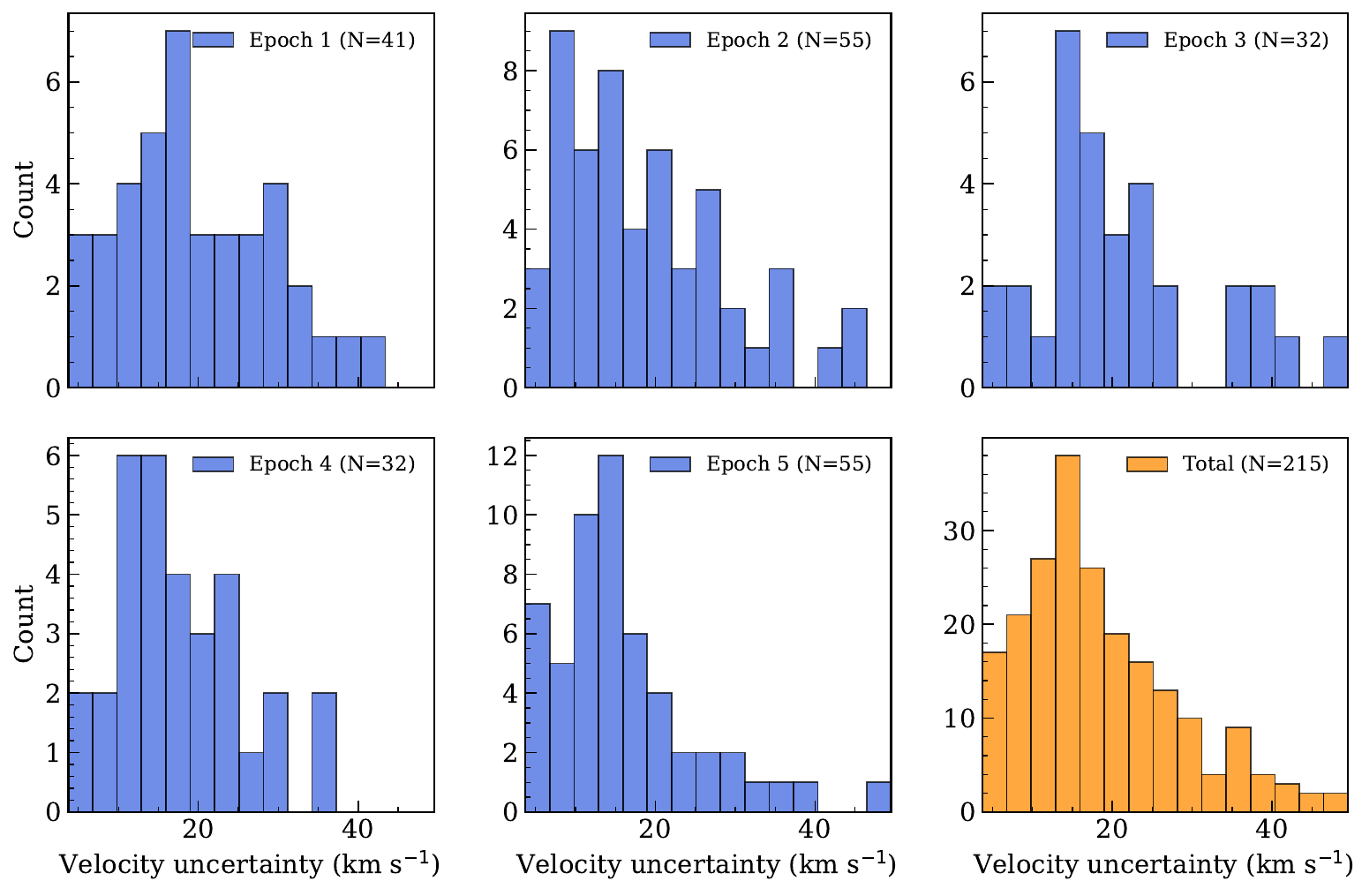}
\caption{Histograms of the velocity measurement uncertainties across the five individual epochs (blue) and for the total combined sample (orange). 
Legends in each panel indicate the number of stars contributing to that epoch. 
The typical uncertainties vary between $\sim 10$ and $\sim 20$~km\,s$^{-1}$ depending on the epoch, with the total distribution reflecting the overall data quality of the Leo~T sample.}
\label{fig:uncertainty_hists}
\end{figure}

As a first step, we examine the distribution of velocity measurement uncertainties across the different epochs.
Figure~\ref{fig:uncertainty_hists} shows histograms of the velocity uncertainties for each of the five epochs, along with the combined distribution of all measurements. 
The typical uncertainties vary between epochs, with the best-quality ones reaching values $\lesssim 10$~km\,s$^{-1}$ and the worst closer to $\sim 20$~km\,s$^{-1}$. 
These distributions also highlight the relative number of stars contributing in each epoch.

In the following Figures \ref{fig:violin_normal} and \ref{fig:violin_limited}, we show the violin plots showcasing the validation test for our methodology. For each of the six different binary fractions tested, from 0 to 1 in steps of 0.2, we simulated 10,000 times the sample observed for Leo T (i.e. 55 stars), following the same methodology as before and fitting for the binary fraction. For these simulations, we used two different binary distribution samples. In the first, shown in Figure \ref{fig:violin_normal}, we used the full binary sample, while limiting the velocity amplitude for the case of Figure \ref{fig:violin_limited}. We see a high degeneracy in the first case, which is due to the presence of binaries with high periods and hence low orbital velocity amplitudes. These have velocity variations from epoch to epoch too small to be detected in the observations. These degeneracies are reduced in the second case, where we limit the minimum velocity amplitude, augmenting the fraction of binaries detected by this method.

\begin{figure}[!h]
\centering
\includegraphics[scale=0.35]{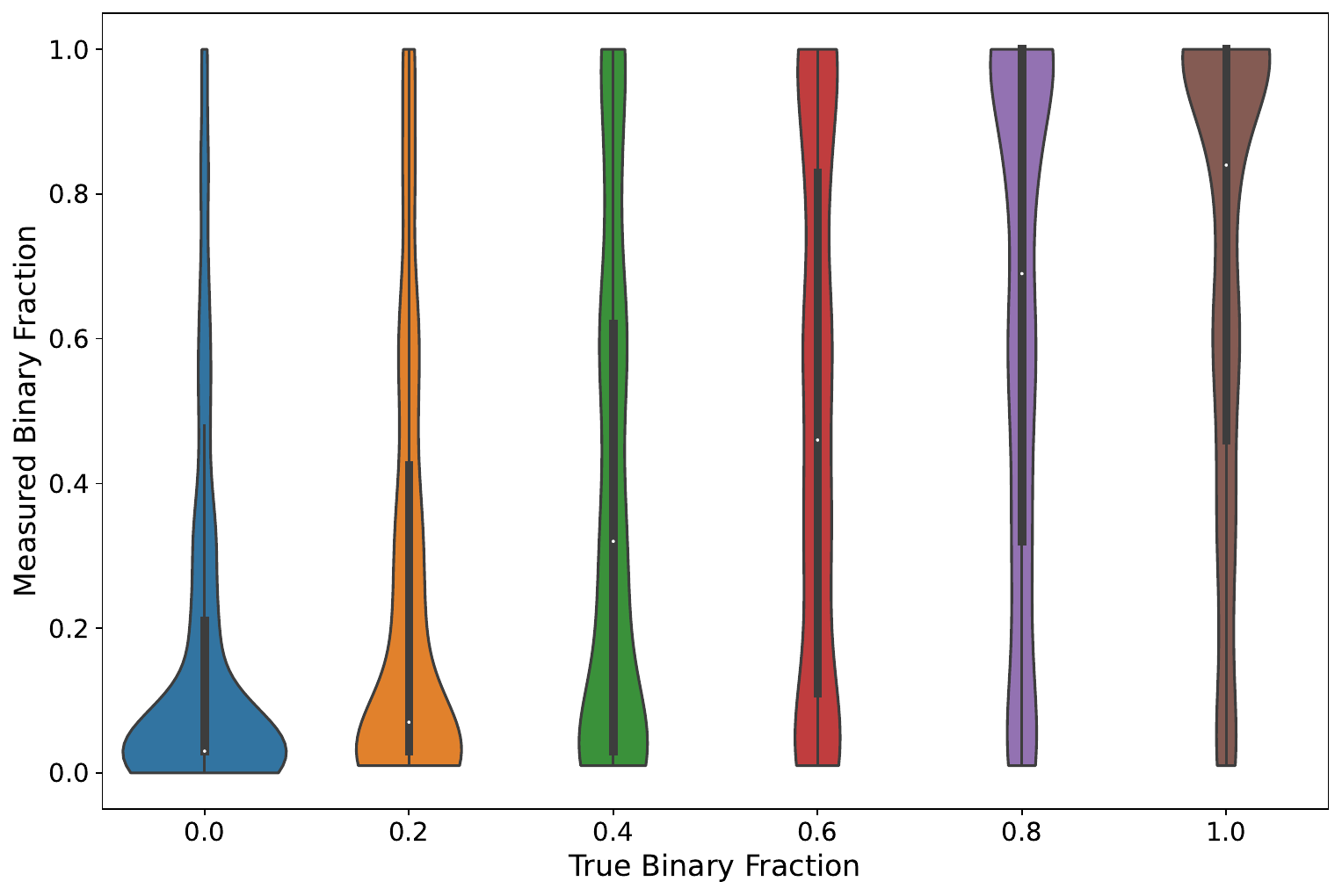}
\caption{Violin plot using the full mock sample. The plot shows how closely our method retrieves the true binary fraction.}
        \label{fig:violin_normal}
\end{figure}

\begin{figure}[!h]
\centering
\includegraphics[scale=0.35]{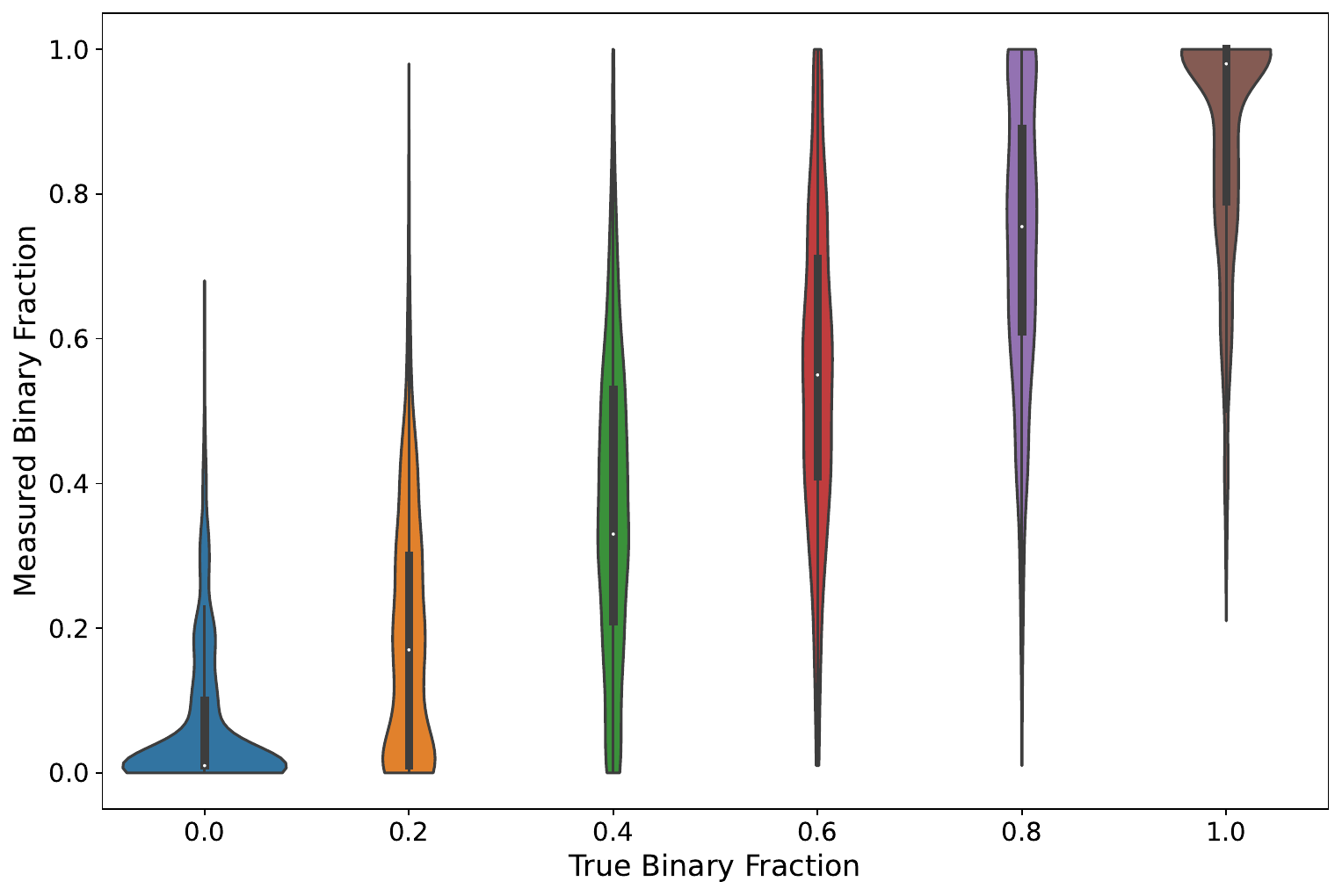}
\caption{Violin plot using the sample limited in semi-amplitude.}
        \label{fig:violin_limited}
\end{figure}

\section{Period-averaged velocities from co-added spectra}\label{apx:test_spexxy}

We constructed a suite of mock spectra to verify that co-adding multi-epoch observations of binary stars yields line-of-sight velocities equal to the average of the combined epochs. Template spectra were drawn from the PHOENIX grid to match typical targets in our sample. For each realisation we assigned a line-of-sight velocity drawn uniformly from $-150$ to $+150\ \mathrm{km\,s^{-1}}$ and applied Gaussian noise to reach a prescribed S/N per pixel. Although several S/N settings were explored, the summary figure below presents the conservative case $\mathrm{S/N}=5$ to reflect our observations (our analysis cut is $\mathrm{S/N}\geq3$, and more than 99\% of spectra have $\mathrm{S/N}>4$). We then formed co-added spectra by combining pairs of epochs with velocities $v_1$ and $v_2$. For each pair we recorded the true mean velocity
\[
\bar{v} = \frac{v_1+v_2}{2},
\]
and the separation $|v_2-v_1|$. Every co-added spectrum was fitted with \spexxy\ using the same configuration as in the main analysis, returning a measured velocity $v_{\rm meas}$. This procedure was repeated for more than 5,000 two-epoch realisations, and analogous tests with three or more epochs behave even more favourably than the two-epoch case discussed here.

\begin{figure}[]
  \centering
  \includegraphics[width=\linewidth]{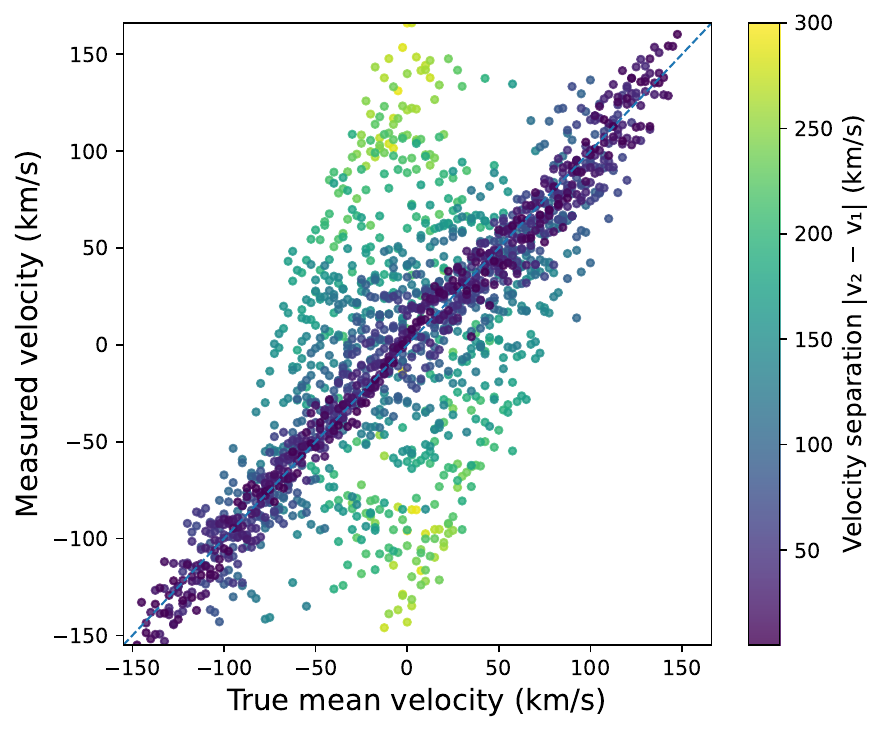}
  \caption{Measured versus true mean velocity for co-added two-epoch mock spectra at $\mathrm{S/N}=5$. Each point represents one realisation. The colour indicates the absolute separation $|v_2-v_1|$. The dashed line marks the 1:1 relation. The strong clustering around the 1:1 line shows that full-spectrum fitting of co-added data recovers the period-averaged velocity of the input epochs. Outliers occur only for very large separations, where the stacked line becomes strongly double peaked.}
  \label{fig:test_peak}
\end{figure}

Co-adding spectra with different velocities broadens the absorption lines, and the full-spectrum fit places the model at the centre of the broadened profile, thereby recovering the mean of the input epochs. In the two-epoch case a clear double peak can appear when $|v_2-v_1|$ is large, but even then the fit typically returns a velocity close to $\bar{v}$. Only for extreme separations, of order $150$–$200\ \mathrm{km\,s^{-1}}$, does the solution occasionally lock onto a single component, which produces outliers. Increasing the number of epochs reduces this failure mode by narrowing the distribution around $\bar{v}$. Figure~\ref{fig:test_peak} compares $v_{\rm mean}$ to $\bar{v}$ for all two-epoch realisations at $\mathrm{S/N}=5$. Points are colour coded by the velocity separation $|v_2-v_1|$, and the tight clustering around the 1:1 line demonstrates that \spexxy\ recovers the period-averaged velocity of co-added spectra, with noticeable deviations confined to the most widely separated cases. 
A closer inspection of these results reveals distinct regimes of stability depending on the velocity separation. For separations $\lesssim 100~km\,s^{-1}$, virtually all realisations follow the 1:1 relation, confirming that the method reliably recovers the mean velocity. Above this threshold, deviations begin to appear, becoming increasingly significant for separations in the range of $150$–$200\ \mathrm{km\,s^{-1}}$. 
Beyond $200$~km\,s$^{-1}$, the fit fails to recover the average velocity. In these extreme cases, rather than converging on the mean (which, for example, would be $\sim 0$~km\,s$^{-1}$ for a pair of spectra at $\pm 150$~km\,s$^{-1}$), \spexxy\ tends to lock onto the velocity of one of the individual components. This behaviour creates the distribution of outliers seen in Figure~\ref{fig:test_peak}, where the measured velocities diverge from the mean and cluster around the values of the individual input velocities. It is important to emphasise, however, that the fraction of successful fits reported by \spexxy\ also decreases substantially in this high-separation regime.

\section{Corner plots}\label{app:corner}

This appendix gathers the full set of corner plots generated by the Markov chain Monte Carlo (MCMC) fits used throughout the paper. In our framework, the observed radial-velocity sample is modelled as a mixture of (i) a Gaussian distribution representing stellar members and (ii) a uniform distribution $U(x)$ representing foreground contaminants. The parameter \texttt{frac} therefore corresponds to the fraction of stars consistent with the Gaussian component, i.e. those identified as galaxy members, while allowing for the presence of outliers due to foreground contamination. Each corner plot displays the posterior probability distributions of the systemic line-of-sight velocity, $v_{\mathrm{los}}$, the intrinsic velocity dispersion, $\sigma_v$, and the fraction \texttt{frac} of stars consistent with the model. 

\begin{figure}[!h]
\centering
\includegraphics[scale=0.45]{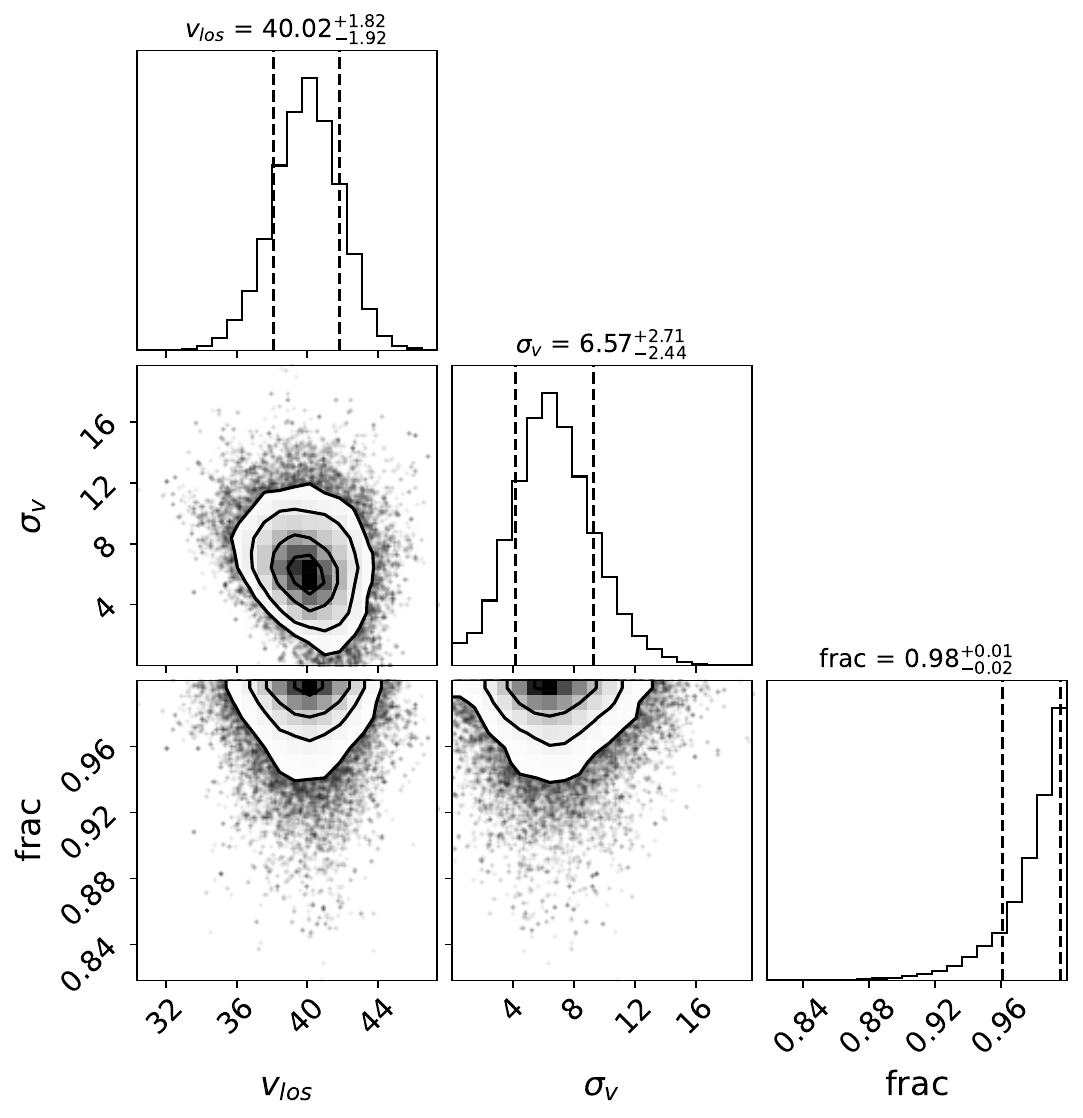}
\caption{Corner plot for the MCMC velocity fit using the entire
  sample of 55 stars. We show the mean value $v_{los}$, dispersion
  $\sigma_v$, and fraction of stars consistent with the model.}
        \label{fig:mcmcvel}
\end{figure}

\begin{figure}[!h]
\centering
\includegraphics[scale=0.45]{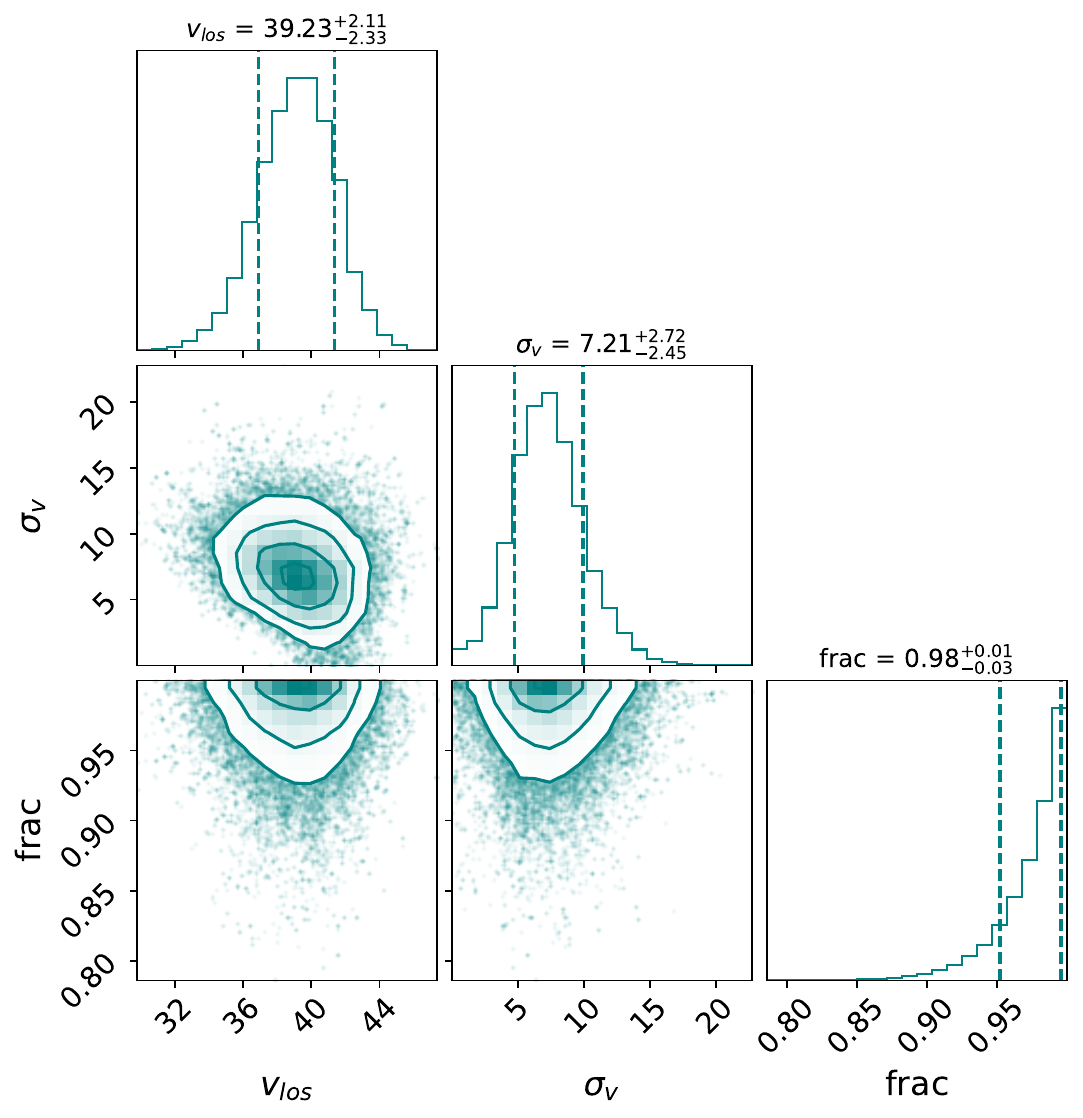}
\caption{Corner plot for the MCMC velocity fit using a
  sample of 44 stars, by removing the 11 stars most likely to be binaries. We show the mean value $v_{los}$, dispersion
  $\sigma_v$, and fraction of stars consistent with the model.}
        \label{fig:mcmcvel}
\end{figure}

\begin{figure}[!h]
\centering
\includegraphics[scale=0.45]{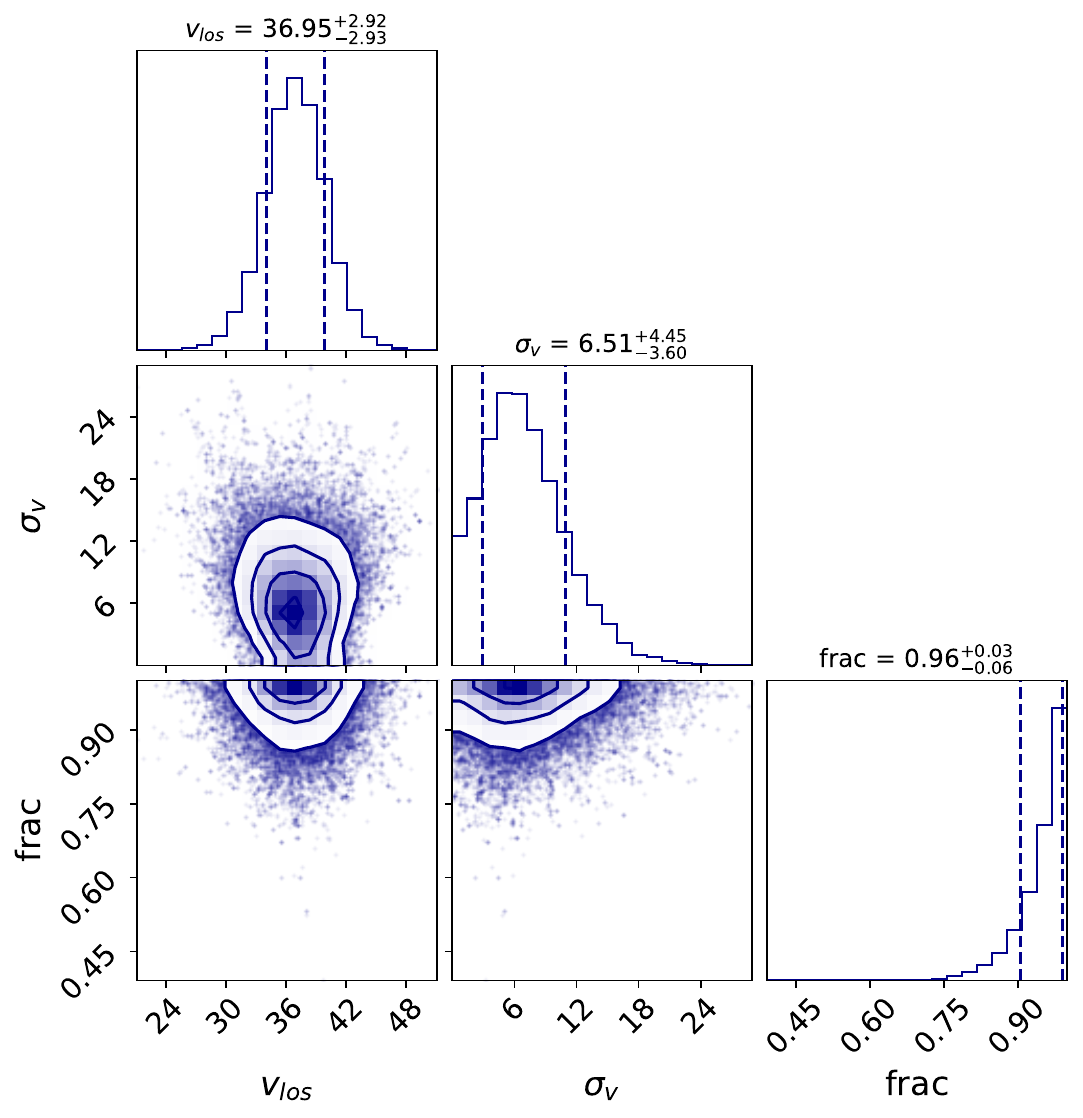}
\caption{Corner plot for the MCMC velocity fit using the entire
  sample of 20 young stars. We show the mean value $v_{los}$, dispersion
  $\sigma_v$, and fraction of stars consistent with the model.}
        \label{fig:mcmcvel}
\end{figure}

\begin{figure}[!h]
\centering
\includegraphics[scale=0.45]{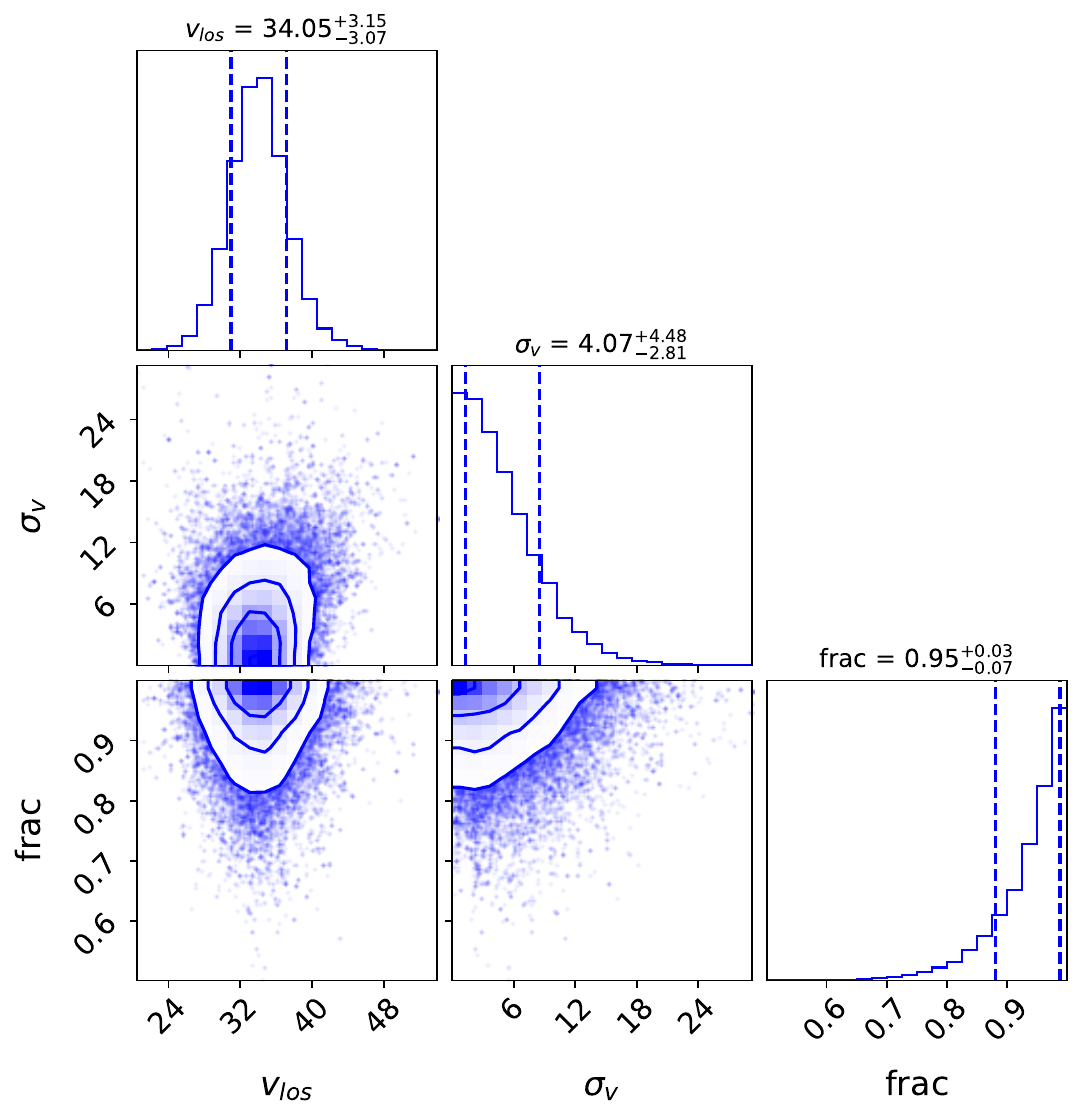}
\caption{Corner plot for the MCMC velocity fit using a
  sample of 14 stars, by removing the six stars most likely to be binaries. We show the mean value $v_{los}$, dispersion
  $\sigma_v$, and fraction of stars consistent with the model.}
        \label{fig:mcmcvel}
\end{figure}

\begin{figure}[!h]
\centering
\includegraphics[scale=0.45]{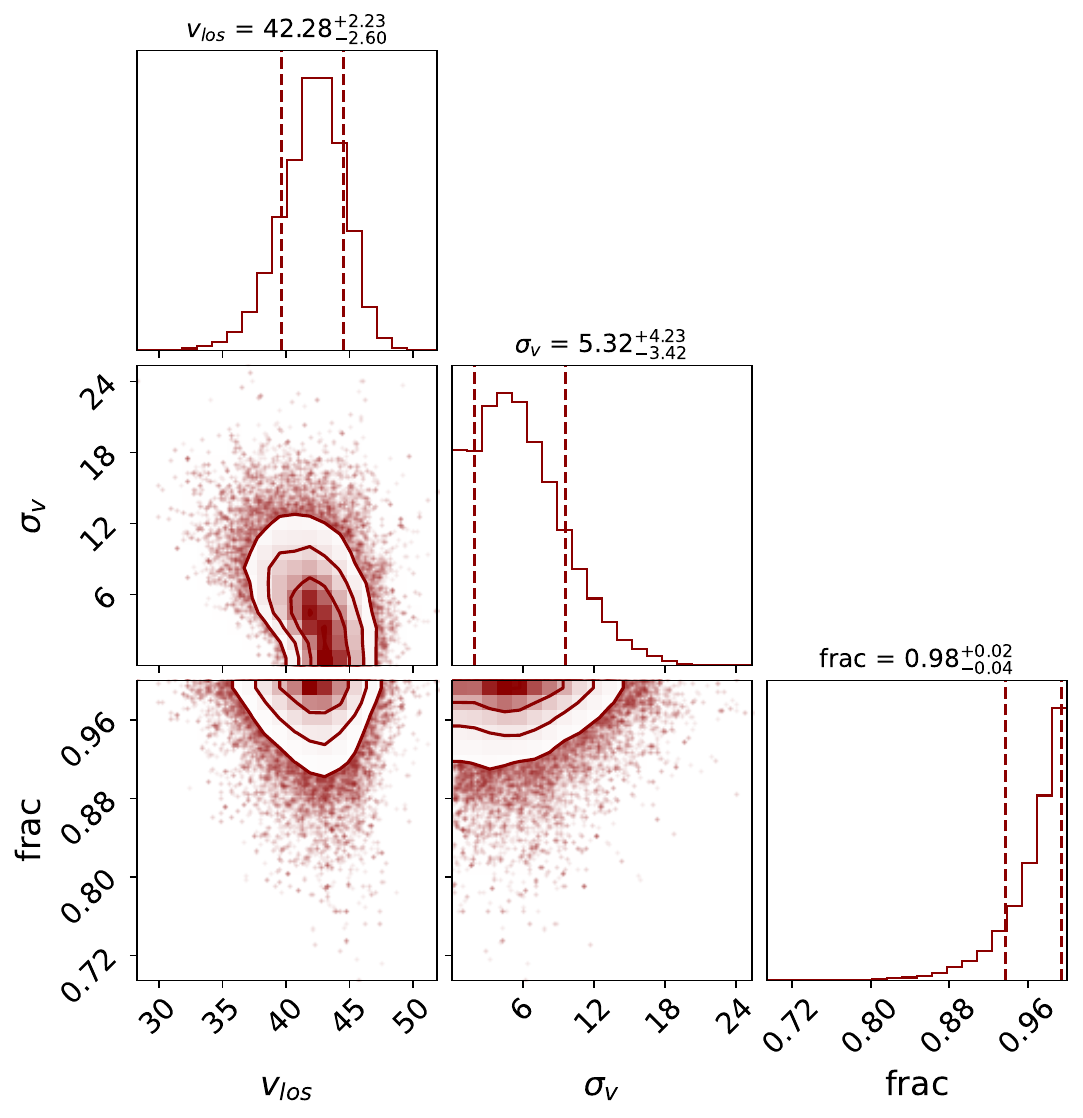}
\caption{Corner plot for the MCMC velocity fit using the entire
  sample of 35 old stars. We show the mean value $v_{los}$, dispersion
  $\sigma_v$, and fraction of stars consistent with the model.}
        \label{fig:mcmcvel}
\end{figure}

\begin{figure}[!h]
\centering
\includegraphics[scale=0.45]{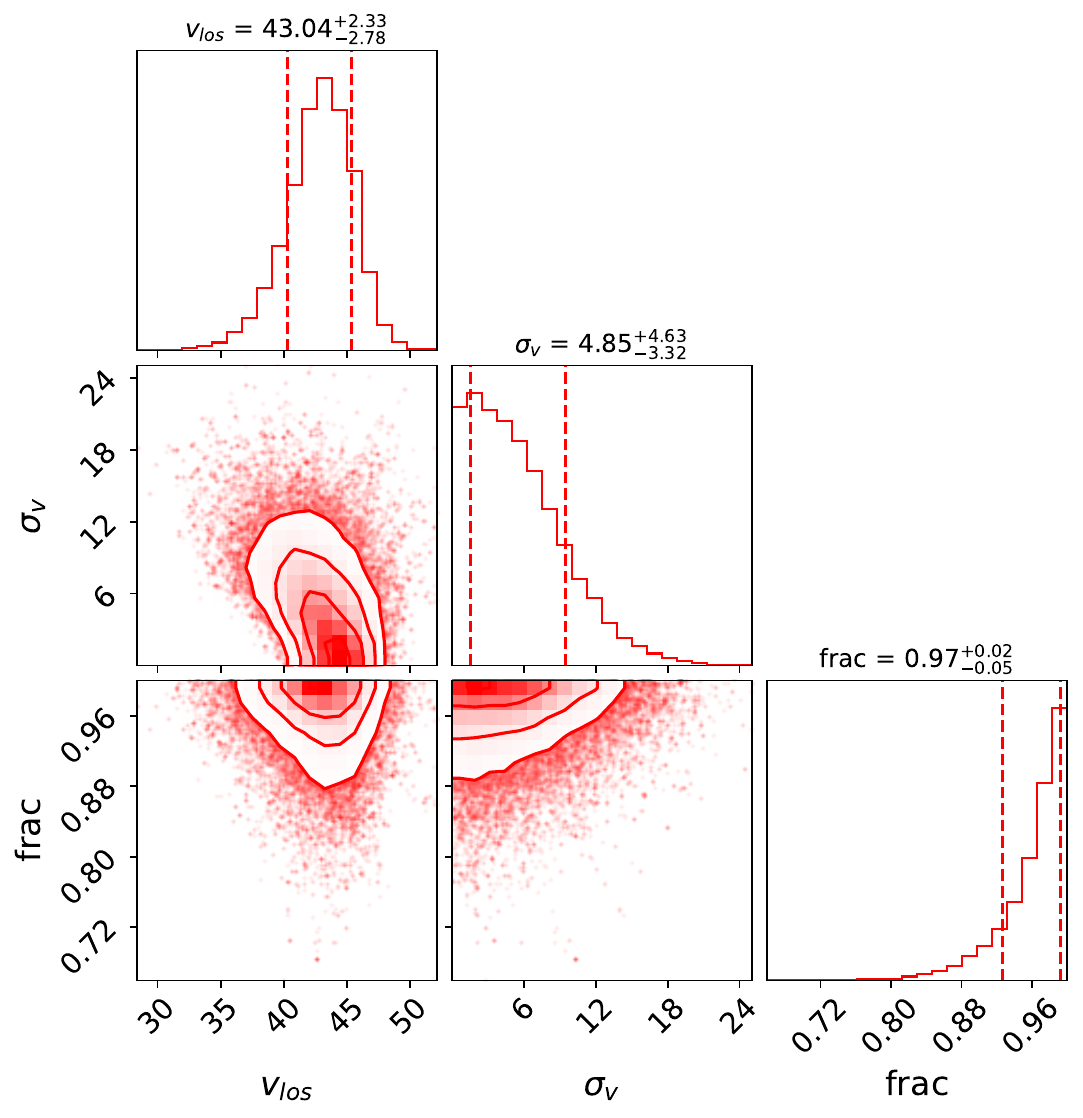}
\caption{Corner plot for the MCMC velocity fit using a
  sample of 32 stars, by removing the three stars most likely to be binaries. We show the mean value $v_{los}$, dispersion
  $\sigma_v$, and fraction of stars consistent with the model.}
        \label{fig:mcmcvel}
\end{figure}

\end{appendix}
\end{document}